\documentclass{article}

\usepackage[T1]{fontenc}
\usepackage{iclr2027_conference,times}
\usepackage{algorithm}
\usepackage{algpseudocode}

\usepackage{amsmath,amsfonts,bm}

\def\eqref#1{equation~\ref{#1}}

\def\1{\bm{1}}

\DeclareMathAlphabet{\mathsfit}{\encodingdefault}{\sfdefault}{m}{sl}
\SetMathAlphabet{\mathsfit}{bold}{\encodingdefault}{\sfdefault}{bx}{n}

\usepackage{hyperref}
\usepackage{url}
\usepackage{graphicx}
\usepackage{booktabs}
\usepackage{array}
\usepackage{colortbl}
\usepackage{placeins}
\definecolor{MainTableOurs}{RGB}{235,243,246}

\title{ReVi: Recursive Recap Video Generation}

\author{
\textbf{Kyrie You Wu\textsuperscript{1}, Chaozhuo Li\textsuperscript{2}, Songyang Liu\textsuperscript{1}, Xueyang Feng\textsuperscript{3}, Ziluowen Luo\textsuperscript{4},} \\
\textbf{Chenxu Wang\textsuperscript{1}, Ji Qi\textsuperscript{1}, Jinfu Zhu\textsuperscript{5}, Litian Zhang\textsuperscript{1}} \\
\textsuperscript{1}Beijing University of Posts and Telecommunications \\
\textsuperscript{2}Beijing Academy of Artificial Intelligence \\
\textsuperscript{3}Renmin University of China \\
\textsuperscript{4}Central South University \\
\textsuperscript{5}Beijing Technology and Business University
}

\iclrfinalcopy

\begin{document}

\maketitle
\fancyhead{}
\renewcommand{\headrulewidth}{0pt}

\begin{abstract}
Novels tell rich stories, but their textual form cannot directly present
the sights and sounds of the worlds they describe. Film adaptations offer
an audiovisual experience but require substantial production effort and
viewing time; film recaps are shorter but depend on existing footage.
Neither route directly provides a concise audiovisual account of a novel.
We therefore study \emph{novel-to-recap} video generation: producing a
self-contained, source-grounded audiovisual retelling directly from a
novel within a fixed viewing-time budget. This task requires compressing
the story while preserving causal connections, allocating sufficient time
to retained events, and aligning narration with newly generated visuals.
We propose \textbf{ReVi}, a dual-loop multi-agent framework guided by
\emph{edit first, generate second}. Its Inner Loop plans story content,
narration time, and corresponding shots before video generation. Across
novels, its Outer Loop performs recursive self-improvement by revising
both the editing policy and the meta policy that governs its revision.
Experiments on two evaluation sets show that ReVi improves recap quality
and duration compliance over representative baselines, with human ratings
and multi-round improvement results providing further support.
\end{abstract}


\section{Introduction}

For centuries, novels have conveyed human experience through rich plots,
complex characters, and imagined worlds, but their textual form cannot directly
present the sights and sounds of those worlds \citep{forster1927aspects}.
Advances in audiovisual technology brought literary narratives to the screen
through film and television adaptations, giving audiences a direct sensory
experience. Producing such adaptations requires substantial resources and
coordinated effort \citep{zuo2026filmworld}, while watching them in full
demands a considerable time commitment. Movie recap videos offer a shorter
way to follow these screen narratives by combining selected clips from an
existing film with voice-over narration that summarizes its main storyline
\citep{shaar2026movierecapsqa}. Unlike the original film, which tells the story through scenes and dialogue, a recap uses narration to summarize key events and explain what happens in scenes it leaves out. 
Clips from the film help illustrate the story. These recaps have become popular on video-sharing platforms such as YouTube~\citep{shaar2026movierecapsqa,ding2025agenticediting}.

Existing agentic work on narrative video production focuses on two tasks: 
\emph{novel-to-film} and \emph{film-to-recap}. 
In \emph{novel-to-film}, multi-agent workflows turn literary narratives into cinematic sequences by developing scripts, storyboards, and shot plans \citep{hu2024storyagent,wu2025movieagent,wang2026mavis}. 
They coordinate visual and audio generation through cinematic planning, continuity management, 
and iterative refinement \citep{huang2026filmaster,huang2026vimax,zuo2026filmworld}. 
In \emph{film-to-recap}, existing work covers different parts of condensing a film or television episode into a short video. Narrative video summarization focuses on identifying key story events and selecting the scenes or shots that depict them \citep{papalampidi2021movie,xie2023narrative,singh2024previously}, 
while narrated-video editing pipelines write summaries, synthesize voice-overs, 
and align narration with footage to assemble a recap \citep{barua2025lotus,ding2025agenticediting,zhou2026videoagent}.

\begin{figure}[t]
    \centering
    \includegraphics[width=\linewidth]{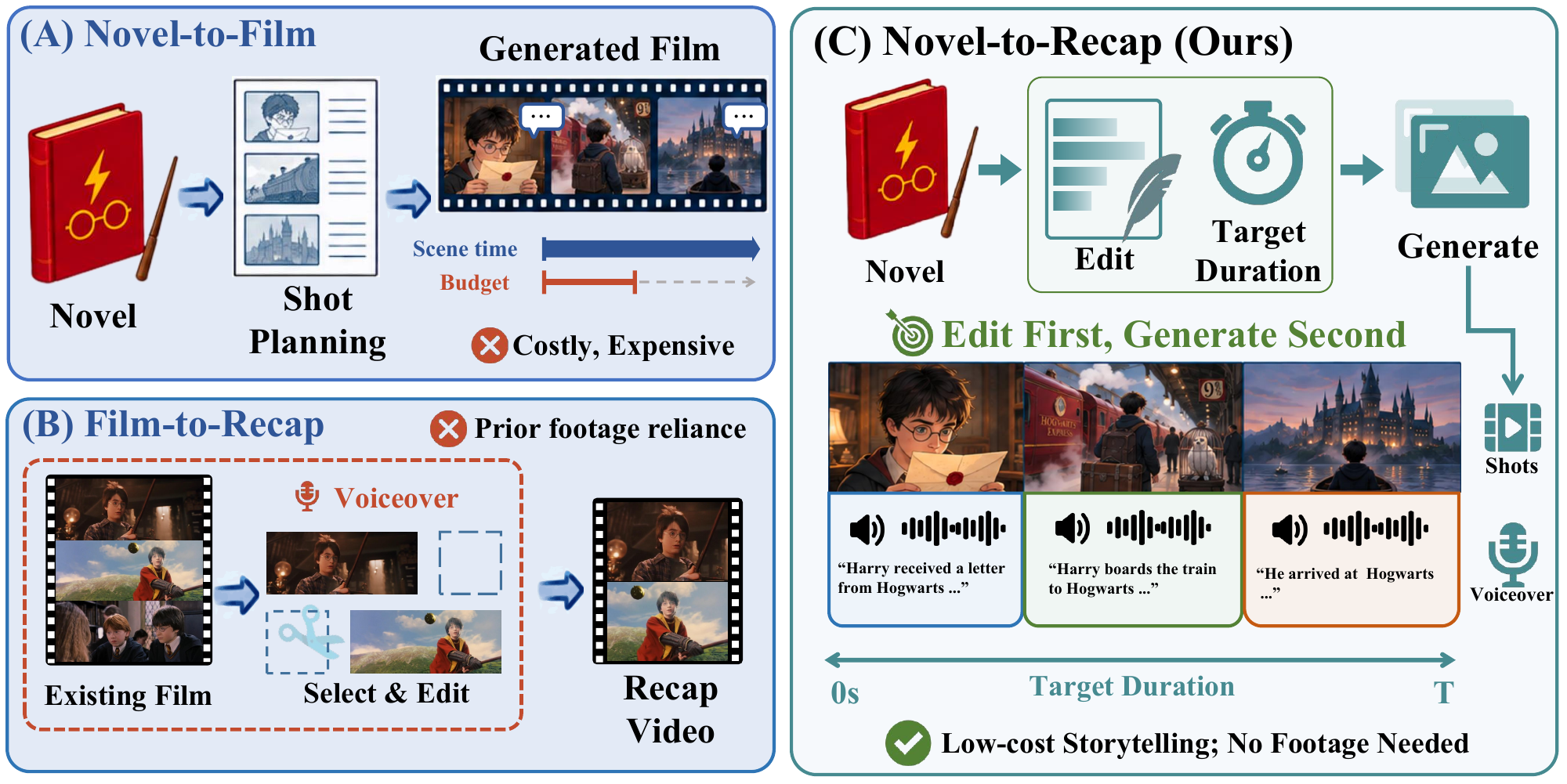}
    \caption{\textbf{From novels to recap videos.} (A) Novel-to-film generates a
    full video adaptation. (B) Film-to-recap selects
    and edits scenes from a source video, combining them with voice-over narration.
    (C) Novel-to-recap produces a narrated audiovisual retelling directly
    from a novel.}
    \label{fig:intro}
\end{figure}

However, neither route directly meets the need to help audiences grasp a
novel's story quickly and intuitively within a fixed viewing-time budget.
As shown in Figure~\ref{fig:intro}, \emph{Novel-to-film} turns literary material into
a full dramatized adaptation, but producing that adaptation is costly and
time-consuming \citep{huang2026vimax,zuo2026filmworld}.
\emph{Film-to-recap} produces an edited video, but it is limited to the scenes and shots available in the source video \citep{ding2025agenticediting,zhou2026videoagent}. This motivates
\emph{novel-to-recap} video generation: producing a self-contained
audiovisual retelling directly from a novel under a fixed viewing-time budget.
Compared with \emph{novel-to-film} and \emph{film-to-recap}, \emph{novel-to-recap}
compresses the storyline through narration and generates only the new shots
needed to visualize it, without requiring either a full dramatized adaptation
or a pre-existing source video.

Realizing this goal poses three challenges. Take the early chapters of
\emph{Harry Potter and the Philosopher's Stone} as an example.
First, \textbf{narrative compression} should preserve the causes, motivations,
and setups needed to understand retained events even after large parts of
the story are omitted. For example, Harry's journey to Hogwarts becomes
difficult to follow if the recap omits the arrival of his Hogwarts letter.
Second, \textbf{temporal allocation} should give retained events enough time
for intelligible narration and perceptible visual action within the fixed
viewing-time budget. A recap may, for instance, compress Harry's childhood
with the Dursleys into a brief introduction while reserving more time for
his arrival at Hogwarts.
Third, \textbf{audiovisual alignment} should coordinate what is narrated
with what is shown, preserving source fidelity and temporal correspondence.
Consider the Sorting Hat scene: when the narration says that Harry is placed
in Gryffindor, the visuals should show the sorting rather than a generic
view of Hogwarts.

To address these challenges, we propose \textbf{ReVi}, a dual-loop multi-agent
framework for \textbf{Re}cursive \textbf{Re}cap \textbf{Vi}deo Generation,
guided by the principle of \emph{edit first, generate second}: determine what
to retain, how much time to allocate, and what to show before generating video
clips. The two loops operate at different scopes. The Inner Loop constructs
a recap video for the current novel, whereas the Outer Loop uses
experience across novels to improve both the reusable editing policy and the
meta policy that governs its revision. 
Within the Inner Loop, the \textbf{Paper Editor} is guided by the Causal
Network Model of narrative comprehension \citep{trabasso1985causal}: it
selects, compresses, omits, or merges source events while preserving their
causal dependencies.
The \textbf{Radio Editor} converts the edited story into timed narration and
allocates the viewing-time budget across retained events, operationalizing
Genette's theory of narrative duration \citep{genette1980narrative}.
The \textbf{AV Planner} organizes the timed narration into Dramatic Beats and
maps them to corresponding shots, adapting Event Segmentation Theory \citep{zacks2007event}.
These editing agents form a pipeline for generating the recap video, progressing from story selection to timed narration and shot planning. 
Across novels, the Outer Loop performs bounded recursive self-improvement by aggregating production and repair traces to
refine both the editing policy and the meta policy governing failure diagnosis
and policy revision. 
Our main contributions are summarized as follows:

\begin{itemize}
    \item To the best of our knowledge, we are the first to formulate and study \emph{novel-to-recap} video generation: the task of producing a
    fixed-duration, source-grounded audiovisual recap directly from a novel
    through concise narration and newly generated visuals.
    \item We propose \textbf{ReVi}, a dual-loop multi-agent framework with
    feedback at two scopes: the Inner Loop plans and repairs each recap, while
    the Outer Loop uses experience across novels to improve both the editing
    policy and the meta policy that governs its revision.
    \item Experiments show that ReVi improves recap quality and duration
    compliance over agentic video-generation baselines.
    Component ablations examine its editing stages, while paired repeated
    searches show separate gains from editing-policy and meta-policy revision.
\end{itemize}


\section{Method}
\label{sec:method}

\begin{figure}[t]
    \centering
    \includegraphics[width=\linewidth]{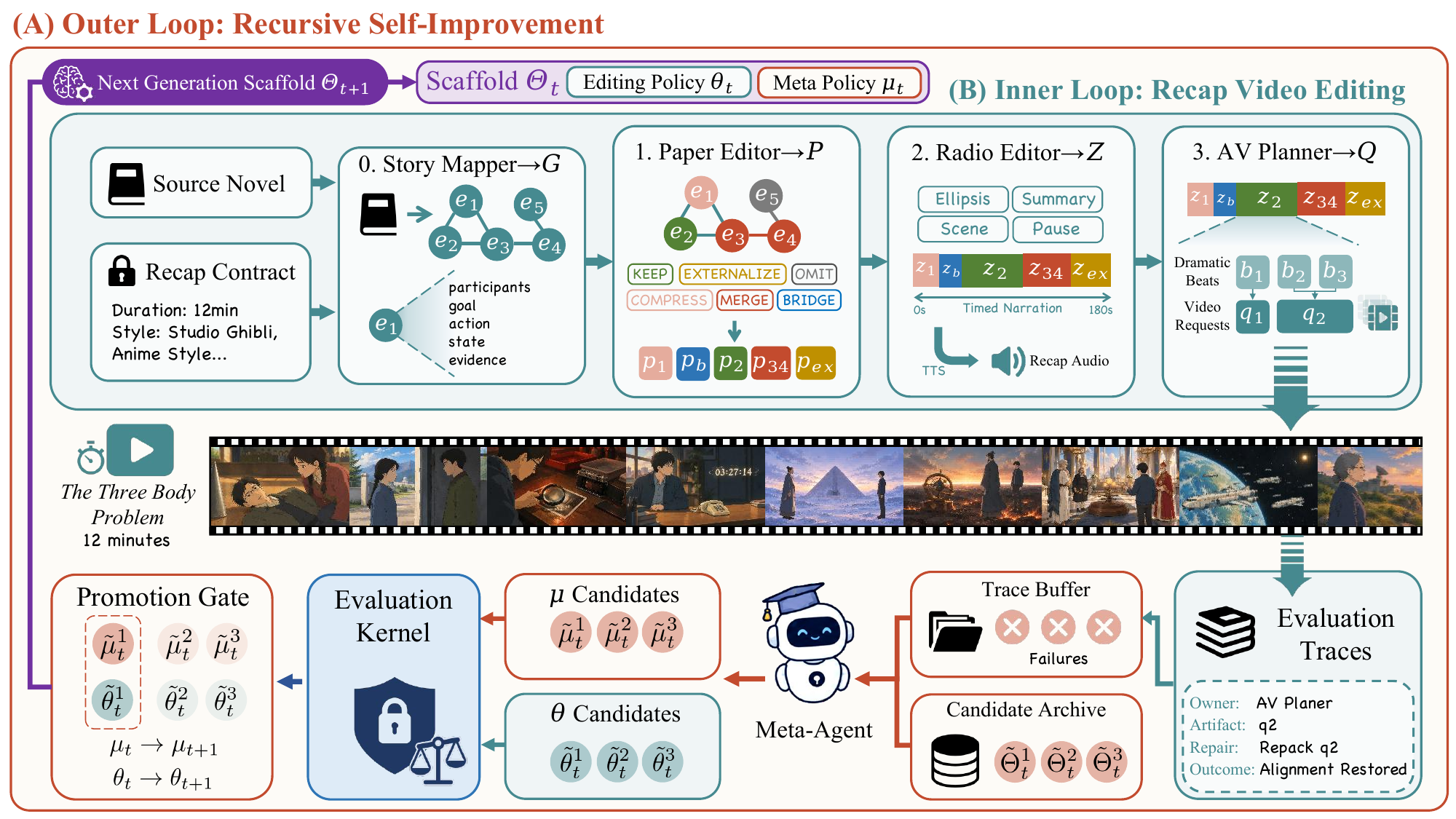}
    \caption{\textbf{Overview of the ReVi framework.} (A) The Outer Loop proposes revisions to the editing
    policy $\theta_t$ and meta policy $\mu_t$ for recursive self-improvement. 
    (B) The Inner Loop passes each recap through the editing agents in sequence before video generation.}
    \label{fig:method}
\end{figure}

Given a source novel $x$ and a recap contract $c$, ReVi produces a
self-contained audiovisual recap $y$. The contract specifies a target duration
$B$, audience, narrative focus, spoiler policy, output language, and visual
style. For example, $c$ may request a 180-second English recap for viewers
unfamiliar with the novel, covering the full plot including the ending, and
rendered in a Studio Ghibli-inspired anime style. As shown in Figure~\ref{fig:method}, ReVi follows two coupled
loops: the Inner Loop constructs
and executes an audiovisual edit for the current novel, while the Outer
Loop uses production experience and improvement history to revise the
reusable policies governing both editing and policy improvement.

\subsection{Inner Loop: Recap Video Generation}
\label{sec:inner_loop}

The Inner Loop follows an \emph{edit first, generate second} workflow for each
novel. Before generating clips, the Paper Editor selects the events needed for
a coherent recap, the Radio Editor writes and times the narration, and the AV
Planner specifies the corresponding on-screen actions and shots. These stages
must be coordinated under the fixed duration budget: retaining an event may
require time to explain its causes, while showing its consequences requires
enough screen time for the action to be understood. The resulting story,
narration, and shot plans guide clip generation and allow coverage or timing
problems to be addressed before footage is produced.

\paragraph{Novel Preprocessing.} A novel distributes story information across long passages and may reveal
events out of chronological order. Therefore, the three editors need a compact, source-verifiable representation that exposes the
dependencies among events before editing begins. The Story Mapper preprocesses the source novel $x$
into a source-linked StoryGraph $G=(V,E)$. First, it
identifies story events as nodes $e_i\in V$, each recording its
participants, motivations, actions, outcomes, and supporting passages. Second,
it adds edges in $E$ for causal dependencies, prerequisites, setup--payoff
relations, and temporal or character-state transitions. Finally, inspired by
Bordwell's three-dimensional model of narrative, which distinguishes the story
world, plot structure, and narration \citep{bordwell2008three}, it records both
the chronological order in which events occur and the disclosure order in
which the novel presents them. The resulting StoryGraph
provides shared evidence for deciding what can be omitted and which
explanations must survive compression. Because every downstream artifact
retains links to the source events and upstream decisions from which it was
derived, an observed failure can be traced through the narration and visual
plan to its editorial origin.

\paragraph{Paper Editor.}
\label{sec:paper_editor}
The StoryGraph exposes the novel's complete event structure, but a
fixed-duration recap cannot present every event. The Paper Editor must
therefore decide what survives compression without removing the causes,
motivations, or setups that make the retained storyline intelligible. Given
the StoryGraph $G$ and recap contract $c$, it produces an ordered
Paper Edit $P$ whose units $p_i$ retain or combine source events, or
supply source-supported bridges and externalized explanations. This process is guided
by the Causal Network Model of narrative
comprehension, in which an event becomes important through its causal
connections and its position in the story's causal chain
\citep{trabasso1985causal}. For example, if a recap retains Harry's arrival at
Hogwarts, it must also retain the arrival of his Hogwarts letter and Hagrid's
revelation that he is a wizard, or replace the omitted material with a brief,
source-supported bridge; otherwise, his journey lacks an explanation.

First, the Paper Editor identifies events central to the requested narrative
focus and traces their causal prerequisites, motivations, and setup--payoff
links in $G$. Second, it assigns \textsc{Keep}, \textsc{Compress},
\textsc{Omit}, or \textsc{Merge} to each event or event chain. If an omission
breaks a necessary connection, \textsc{Bridge} inserts a concise,
source-supported explanation; if understanding depends on an internal state,
\textsc{Externalize} assigns that information to narration. Finally, a fixed
closure checker verifies source support, prerequisite coverage, setup--payoff
relations, and reveal order, returning any broken dependency for revision.
Once these checks pass, $P$ records the selected event sequence, its source
links, and the operation applied to each unit. This gives the Radio Editor a
compact, causally coherent account to schedule, preventing missing
explanations from surfacing only after timing or video generation.
Appendix~\ref{sec:method_details} defines the operations.

\paragraph{Radio Editor.}
\label{sec:radio_editor}
The Paper Edit determines which events survive compression, but it does not
determine how much narration time each event should receive. Under a fixed
viewing-time budget, treating all retained events equally would either exceed
the budget or compress important developments until they become difficult to
follow. Given the Paper Edit $P$ and recap contract $c$, the Radio Editor
produces timed narration $Z$ as source-linked segments $z_i$ anchored
to the relevant edit units in $P$. This process is grounded in
Genette's theory of
narrative duration, which distinguishes \emph{ellipsis, summary, scene, and
pause} according to the relation between story time and narrative time
\citep{genette1980narrative}. For example, a recap of the first \emph{Harry
Potter} novel may use ellipsis to skip an uneventful interval, summarize
Harry's childhood with the Dursleys, present the Sorting Ceremony as a scene,
and pause briefly to explain the Hogwarts houses.

First, the Radio Editor assigns a narration mode to each unit in $P$ and
allocates the available time according to its narrative function and
information density. Here, \textsc{Scene} denotes comparatively detailed
treatment rather than literal equality between story time and playback time.
Second, it writes the narration, with each segment recording its source events,
introduced facts, assigned mode, and text. Finally, text-to-speech converts the
draft into measured timestamps and checks whether its total duration satisfies
the budget $B$. If the narration is overlong, the Radio Editor first removes
redundant wording; persistent overload is returned to the Paper Editor for
structural compression. If time remains, it expands necessary context or gives
key events fuller treatment. The resulting $Z$ is a source-linked,
budget-compliant narration with measured timestamps, providing the AV Planner
with a stable temporal backbone for shot planning.

\paragraph{Audio-Visual Planner.}
\label{sec:av_planner}
Timed narration specifies what is said and when, but its sentence boundaries
do not necessarily correspond to perceptible events or feasible shots. The AV
Planner must therefore decide how each narrated development should unfold on
screen while keeping the visuals synchronized with the narration. Given the
StoryGraph $G$, timed narration $Z$, and recap contract $c$, it
produces an audiovisual plan $Q$. This process is grounded in Event
Segmentation Theory, which holds that viewers maintain a representation of the
current event and perceive a boundary when prediction error triggers an update
to that representation \citep{zacks2007event}. For example, a single sentence
may narrate the Sorting Hat considering Slytherin, announcing Gryffindor, and
the hall erupting in applause. Each change in action or outcome updates the
viewer's understanding of the event, so the planner represents these moments
as separate \emph{Dramatic Beats}.

First, the AV Planner segments $Z$ at perceptible changes in goals, actions,
outcomes, character focus, time, or location, forming dramatic beats $b$.
Each Beat records its source
events, initial and outcome states, narration anchor, screen interval, and
continuity group. Second, the planner determines the division of information
between modalities: narration carries internal states, background, and elapsed
time, while visuals depict selected observable actions and state changes. It
then maps each Beat to one or more shot intents. Finally, it groups compatible
shots into timestamped generation requests $q$ with visual references, checking
compatibility in characters, setting, state transitions, timing, and
continuity. If an action cannot be presented intelligibly within its assigned
interval, the planner requests revised timing or upstream compression. The
resulting $Q$ provides a source-grounded, temporally aligned blueprint that
guides a video generation model in producing clips.

\paragraph{Repair and Trace Collection.}
After clip generation and assembly, ReVi repairs the current recap and records
each production attempt in a trace buffer $\mathcal{T}$. First, a fixed temporal
critic checks event coverage, event order, narration--visual alignment, and
compliance with the target runtime. Second, each detected failure is converted
into a repair request containing the supporting evidence and affected
artifact. The orchestrator routes this request to the earliest responsible
stage: source reconstruction, content selection, temporal allocation, or
visual planning. Third, the responsible stage revises its artifact under the
fixed editing policy $\theta_t$, after which dependent outputs are recomputed
and affected clips are regenerated while unaffected artifacts are reused. This
cycle continues until all checks pass or the repair budget is exhausted.
Finally, ReVi records the decisions, intermediate artifacts, critic feedback,
repairs, and outcomes from every attempt as a production trace. The repairs
update only the current recap; the collected traces provide evidence for the
Outer Loop.

\subsection{Outer Loop: Recursive Self-Improvement}
\label{sec:outer_loop}

The Inner Loop's production traces feed the Outer Loop. At generation $t$,
ReVi maintains the system state in Equation~\eqref{eq:system_state}:
\begin{equation}
    \Theta_t=(\theta_t,\mu_t),
    \label{eq:system_state}
\end{equation}
where $\theta_t$ is the editing policy and $\mu_t$ guides its improvement.
Both consist of versioned prompt instructions, examples, and configuration rules.
First, the Meta-Agent uses the trace buffer $\mathcal{T}$ and archive
$\mathcal{A}$ of prior revision attempts to propose candidate states
$\widetilde{\Theta}_t^{(i)}$, each changing one policy at a time. Second,
the frozen Evaluation Kernel $E^*$ supplies scores for the fixed promotion
rules. Finally, ReVi records each attempt and its outcome in $\mathcal{A}$;
accepted revisions form the next state $\Theta_{t+1}$.

\paragraph{Editing Policy.}
\label{sec:policy_representations}
The editing policy in Equation~\eqref{eq:editing_policy} is
\begin{equation}
    \theta_t=(\pi_t^{S},\pi_t^{P},\pi_t^{R},\pi_t^{AV},\rho_t),
    \label{eq:editing_policy}
\end{equation}
where the first component governs source interpretation, the next three
govern the editing agents, and $\rho_t$ governs coordination and repair.
Editable rules determine which motivations and dependencies to recover,
which events to retain or compress, how to assign narration modes and time,
how to define and group Beats, and when to request upstream revision.
A patch may strengthen the condition for retaining a payoff's setup,
change how explanatory passages receive time, or revise when dense Beats
are split. Such changes persist across novels. Evolution revises how the editors apply them.

\paragraph{Editing-Policy Evolution.}
\label{sec:editor_search}
The Meta-Agent aggregates traces across novels and diagnoses recurring
failures by inspecting source evidence, intermediate edits, and attempted
repairs. It proposes editing-policy candidates
$\widetilde{\theta}_t^{(i)}$, each specifying a target rule, triggering
condition, revised instruction or configuration, expected effect, and
constraints. An alignment failure, for example, can originate in visual
realization, insufficient screen time, or excessive narration density.
This use of execution feedback follows reflective
prompt optimization \citep{agrawal2026gepa}, with revisions restricted to
ReVi's editable policies. Candidates undergo structural and narration
checks before expensive rendering; changes affecting visual coverage or
timing are evaluated through generated outputs. The fixed kernel assesses
source support, causal closure, comprehension, duration, audiovisual
alignment, and cost. Only candidates satisfying its acceptance constraints
are eligible for promotion under a fixed quality--cost rule and margin.
The selected editor becomes $\theta_{t+1}$; if none passes, then
$\theta_{t+1}=\theta_t$. Every attempt is archived with its
diagnosis, patch, outcome, regressions, and cost, so later searches can
examine whether an earlier intervention supported its diagnosis.

\paragraph{Meta Policy.}
The meta policy $\mu_t$ determines how to diagnose failures, propose patches,
and allocate a fixed search budget. Its updates address recurring problems
in improvement itself: incorrect attribution, ineffective revisions, or
repeated spending on unproductive candidates. A diagnosis rule may require
checking missing causal information before blaming rendering; a proposal
rule may favor conditional instructions over generic prompt expansion; an
allocation rule may favor testing alternative Paper Edits over repeated
video generation. A new diagnosis for one failure is an output of
$\mu_t$; changing the reusable diagnostic procedure updates $\mu_t$ itself.
The current meta policy uses production traces and archived improvement
attempts to propose these revisions, following self-referential improvement
approaches \citep{fernando2024promptbreeder,zelikman2024stop}.

\paragraph{Meta-Policy Evaluation.}
\label{sec:meta_policy_improvement}
To evaluate a candidate meta policy $\widetilde{\mu}_t^{(j)}$, ReVi compares
it with $\mu_t$ under matched inputs and search budget; its index $j$ is
independent of the editing-policy candidate index $i$.
First, each comparison episode $h\in\mathcal{E}_{\mathrm{meta}}$ fixes a parent
editor $\theta_h$, production traces $\mathcal{T}_h$, improvement history
$\mathcal{A}_h$, and search budget $K$. Holding the tested meta policy $\mu$
fixed within the episode, the search selects an editor using development and
promotion data under the fixed acceptance procedure:
\begin{equation}
    \widehat{\theta}_h(\mu)=\operatorname{Search}_{\mu}
    (\theta_h,\mathcal{T}_h,\mathcal{A}_h;K).
    \label{eq:editor_search}
\end{equation}
Second, the editor $\widehat{\theta}_h(\mu)$ selected in
Equation~\eqref{eq:editor_search} is frozen and evaluated on the episode's
held-out novels $\mathcal{D}^{\mathrm{held}}_h$. Writing
$J_{E^*}$ for the frozen kernel's utility relative to the parent editor
$\theta_h$, ReVi averages these scores across episodes:
\begin{equation}
    V(\mu)=\frac{1}{|\mathcal{E}_{\mathrm{meta}}|}
    \sum_{h\in\mathcal{E}_{\mathrm{meta}}}
    J_{E^*}\!\left(\widehat{\theta}_h(\mu);
        \theta_h,\mathcal{D}^{\mathrm{held}}_h\right).
    \label{eq:outer_meta}
\end{equation}
Thus Equation~\eqref{eq:outer_meta} evaluates a search procedure through the
downstream improvement of the editors it produces, as in Hyperagents
\citep{zhang2026hyperagents}.
Finally, the promotion gate compares the candidate with $\mu_t$ using this
score. The candidate becomes $\mu_{t+1}$ only if it passes the fixed promotion
criterion without increasing permitted constraint violations; otherwise
$\mu_{t+1}=\mu_t$. Together with the accepted editor $\theta_{t+1}$, it forms
$\Theta_{t+1}=(\theta_{t+1},\mu_{t+1})$, which becomes the current scaffold
in the next generation. Its accepted meta policy governs subsequent editor
searches and further revisions to the improvement procedure. Task contracts,
evaluation definitions, and final
acceptance rules remain fixed, with a sealed test partition reserved for
assessing the resulting system. If a proposed editor revision fails
acceptance, $\theta_{t+1}=\theta_t$.


\section{Experiments}
\label{sec:experiments}

\subsection{Experimental Settings}
\label{sec:experimental_setup}
\paragraph{Datasets and duration contracts.}
We evaluate recap generation across difficulty-graded narratives and
longer novels using FilmEval and Recap32. \textbf{FilmEval}~\citep{zuo2026filmworld}
provides source narratives grouped into Easy, Medium, and Hard tiers by
narrative complexity. We use these source texts as recap inputs and
assign a duration target to each tier. \textbf{Recap32} draws on longer
English Gutenberg novels from NarrativeQA~\citep{kocisky2018narrativeqa}.
The protocol assigns separate novels to policy evolution, editor
promotion, meta-policy evaluation, and testing. We analyze the 12
designated test novels under a 180-second duration target.

\paragraph{Metrics.}
Three Gemini~3.7 Flash judgments score each completed video on three quality dimensions.
\textbf{Recap Fidelity (RF)} combines weighted coverage of essential
source events, the logical consistency of their relationships, and
resistance to unsupported content, using a shared source reference.
\textbf{Narrative Pacing (NP)} scores how clearly the finished recap
develops its story, including plot progression, transitions, and the
distribution of information over playback time.
\textbf{Audio-Visual Grounding (AVG)} measures the share of expected
narration segments whose visuals support or complement the corresponding
spoken content. We assess temporal control separately:
\textbf{Duration Error (DE)} is the absolute difference between final and
target playback times, expressed as a percentage of the target, and
\textbf{Duration Pass Rate (DP)} is the share of evaluated outputs that
both complete and satisfy the prescribed duration tolerance. Keeping
timing separate from quality makes coverage gains interpretable under
the same viewing-time contract.

\paragraph{Baselines and scoring.}
ViMax~\citep{huang2026vimax} is the primary system-level baseline.
The FilmEval comparison also includes VideoClaw~\citep{hitsztmg2026videoclaw},
MovieAgent~\citep{wu2025movieagent},
MM-StoryAgent~\citep{xu2025mmstoryagent}, and
VGoT~\citep{zheng2025videogenofthought}. Direct LLM creates narration
and a storyboard in a single pass before common rendering. Recap32
compares Direct LLM, ViMax, MovieAgent, and ReVi. All methods receive the same
output request; baselines receive the target duration only in the prompt.
The appendix gives more details.

\subsection{Main Comparison}
\label{sec:main_results}
\paragraph{Recap quality and duration compliance.}
Table~\ref{tab:main_results} compares ReVi with Direct LLM and agentic
video-generation baselines on FilmEval and Recap32. Each method follows
the same recap task and output specification, and we assess completed
videos for Recap Fidelity, Narrative Pacing, Audio-Visual Grounding, and Duration
Compliance. ReVi leads on the quality dimensions across both
collections. The strongest baseline varies by dimension, indicating
that plot accuracy, intelligibility, and visual correspondence place
distinct demands on recap generation. ReVi's joint gains are
consistent with its connected sequence of causal selection, timed
narration, and shot planning. The appendix reports a shared-narration
control (Table~\ref{tab:equal_duration}) and component interventions.

\begin{table}[t]
\centering
\caption{\textbf{Recap quality and duration compliance.} RF/NP/AVG are quality scores; DE/DP are percentages (DP pass counts in parentheses). Best and second-best are bold and underlined.}
\label{tab:main_results}
\vspace{3pt}
\begingroup
\normalsize
\setlength{\tabcolsep}{0pt}
\setlength{\arrayrulewidth}{0.3pt}
\renewcommand{\arraystretch}{1.00}
\setlength{\aboverulesep}{1pt}
\setlength{\belowrulesep}{1pt}
\arrayrulecolor{black}
\begin{tabular}{@{}>{\raggedright\arraybackslash\hspace{2pt}}p{0.230\linewidth}|>{\centering\arraybackslash}p{0.140\linewidth}>{\centering\arraybackslash}p{0.140\linewidth}>{\centering\arraybackslash}p{0.140\linewidth}|>{\centering\arraybackslash}p{0.130\linewidth}>{\centering\arraybackslash}p{0.210\linewidth}@{}}
\toprule[0.8pt]
Method & RF $\uparrow$ & NP $\uparrow$ & AVG $\uparrow$ & DE $\downarrow$ & DP $\uparrow$ \\
\specialrule{0.4pt}{1.5pt}{1.2pt}
\multicolumn{6}{c}{\textit{FilmEval: 15 sources; 90/120/150 s}} \\
\midrule[0.4pt]
Direct LLM & 74.46 & 66.51 & 61.62 & 226.70 & 0.0\% (0/15) \\
ViMax & 82.09 & 78.44 & \underline{78.66} & \underline{108.71} & 0.0\% (0/15) \\
VideoClaw & 81.11 & \underline{79.20} & 77.30 & 145.24 & 0.0\% (0/15) \\
MovieAgent & \underline{83.72} & 78.61 & 74.73 & 160.34 & 0.0\% (0/15) \\
MM-StoryAgent & 75.57 & 73.77 & 69.90 & 219.40 & 0.0\% (0/15) \\
VGoT & 77.48 & 76.20 & 72.65 & 163.53 & 0.0\% (0/15) \\
\rowcolor{MainTableOurs}
\textbf{ReVi} & \textbf{91.58} & \textbf{90.20} & \textbf{88.67} & \textbf{2.37} & \textbf{100.0\% (15/15)} \\
\specialrule{0.4pt}{1pt}{1pt}
\multicolumn{6}{c}{\textit{Recap32: 12 test novels; 180 s}} \\
\midrule[0.4pt]
Direct LLM & 68.95 & 66.40 & 63.25 & 215.72 & 0.0\% (0/12) \\
ViMax & 78.30 & 76.35 & \underline{74.07} & \underline{125.20} & 0.0\% (0/12) \\
MovieAgent & \underline{80.99} & \underline{77.50} & 72.30 & 191.31 & 0.0\% (0/12) \\
\rowcolor{MainTableOurs}
\textbf{ReVi} & \textbf{90.68} & \textbf{88.25} & \textbf{82.88} & \textbf{2.07} & \textbf{100.0\% (12/12)} \\
\specialrule{0.4pt}{1.5pt}{1.2pt}
\specialrule{0.8pt}{0pt}{0pt}
\end{tabular}
\endgroup
\end{table}

\begin{table}[t]
\centering
\caption{\textbf{Human evaluation of recap quality.} Mean $\pm$ SD of individual FilmEval ratings (0--5). Best and second-best means are bold and underlined.}
\label{tab:human_results}
\vspace{3pt}
\begingroup
\normalsize
\setlength{\tabcolsep}{0pt}
\setlength{\arrayrulewidth}{0.3pt}
\renewcommand{\arraystretch}{1.00}
\setlength{\aboverulesep}{1pt}
\setlength{\belowrulesep}{1pt}
\arrayrulecolor{black}
\begin{tabular}{@{}>{\raggedright\arraybackslash\hspace{2pt}}p{0.220\linewidth}|>{\centering\arraybackslash}p{0.255\linewidth}>{\centering\arraybackslash}p{0.255\linewidth}>{\centering\arraybackslash}p{0.255\linewidth}@{}}
\toprule[0.8pt]
Method & RF $\uparrow$ & NP $\uparrow$ & AVG $\uparrow$ \\
\specialrule{0.4pt}{1.5pt}{1.2pt}
Direct LLM & 3.44 $\pm$ 0.51 & 3.00 $\pm$ 0.34 & 3.06 $\pm$ 0.42 \\
ViMax & \underline{4.17} $\pm$ 0.51 & \underline{3.94} $\pm$ 0.64 & \underline{4.00} $\pm$ 0.49 \\
VideoClaw & 3.78 $\pm$ 0.43 & 3.83 $\pm$ 0.62 & 3.78 $\pm$ 0.43 \\
MovieAgent & 4.06 $\pm$ 0.42 & 3.50 $\pm$ 0.62 & 3.39 $\pm$ 0.50 \\
MM-StoryAgent & 3.33 $\pm$ 0.49 & 3.44 $\pm$ 0.51 & 2.94 $\pm$ 0.42 \\
VGoT & 3.67 $\pm$ 0.49 & 3.67 $\pm$ 0.49 & 3.56 $\pm$ 0.51 \\
\rowcolor{MainTableOurs}
\textbf{ReVi} & \textbf{4.44} $\pm$ 0.51 & \textbf{4.56} $\pm$ 0.51 & \textbf{4.50} $\pm$ 0.62 \\
\specialrule{0.4pt}{1.5pt}{1.2pt}
\specialrule{0.8pt}{0pt}{0pt}
\end{tabular}
\endgroup
\end{table}

\paragraph{Duration under a fixed viewing time.}
To assess duration control, we compare each completed video's playback
length with its assigned target and report DE and DP alongside the
quality scores in Table~\ref{tab:main_results}. ReVi meets the requested
duration in the supplied comparisons and achieves higher quality with
substantially shorter playback than the baselines.
Longer outputs offer additional room to mention events and extend scenes;
ReVi's joint quality and duration results point to efficient allocation
of the available viewing time. This pattern is consistent with the Radio
Editor's allocation of narration time before rendering and its use of
measured speech timing to refine the plan. Under a fixed budget, causal
links require deliberate selection and enough narration time to remain
intelligible.

\subsection{Human Evaluation}
\label{sec:human_evaluation}
To test whether the automatic quality gains carry over to human viewing,
we ask raters to evaluate source-matched FilmEval recaps from ReVi and
the evaluated baselines. Every method is judged on the same source set.
Raters have access to the source text and shared event reference, while
method identities are hidden and presentation order is randomized.
They score recap fidelity, narrative pacing, and audio-visual grounding
on a common scale; Table~\ref{tab:human_results} summarizes their
ratings across methods. ReVi receives the highest ratings in each
dimension, with the most pronounced separation in narrative pacing.
This pattern is consistent with the editing pipeline: causal event
selection preserves the context needed to understand later developments,
timed narration regulates how that context unfolds, and beat-aligned
shots reinforce the spoken account. The joint preference across
dimensions supports the interpretation that the automatic gains reflect
qualities perceptible in the finished recaps. The appendix details the
rating aggregation, source-paired comparisons, and alignment between
automatic scores and human judgments.

\begin{figure}[t]
\centering
\includegraphics[width=\linewidth]{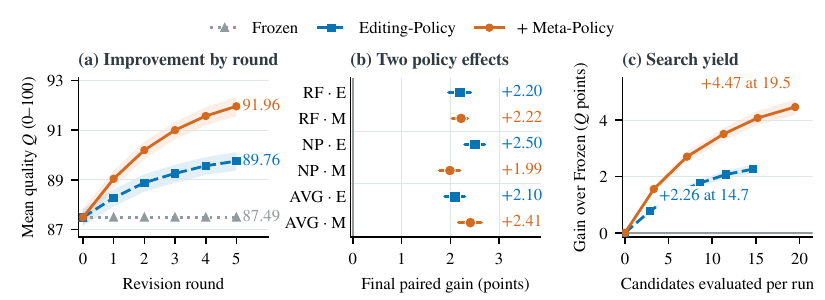}
\caption{\textbf{Outer Loop evaluation.}
Ten paired searches are compared over five revision rounds.
(a) Mean quality $Q=(\mathrm{RF}+\mathrm{NP}+\mathrm{AVG})/3$ for Frozen,
Editing-Policy, and Editing-Policy + Meta-Policy.
(b) Final run-paired differences for editing-policy updates
(E: Editing-Policy minus Frozen) and added meta-policy updates
(M: Editing-Policy + Meta-Policy minus Editing-Policy).
(c) Mean $Q$ gain over Frozen versus cumulative candidates evaluated
per run.}
\label{fig:outer_loop}
\end{figure}

\begin{figure}[t]
\centering
\includegraphics[width=\linewidth]{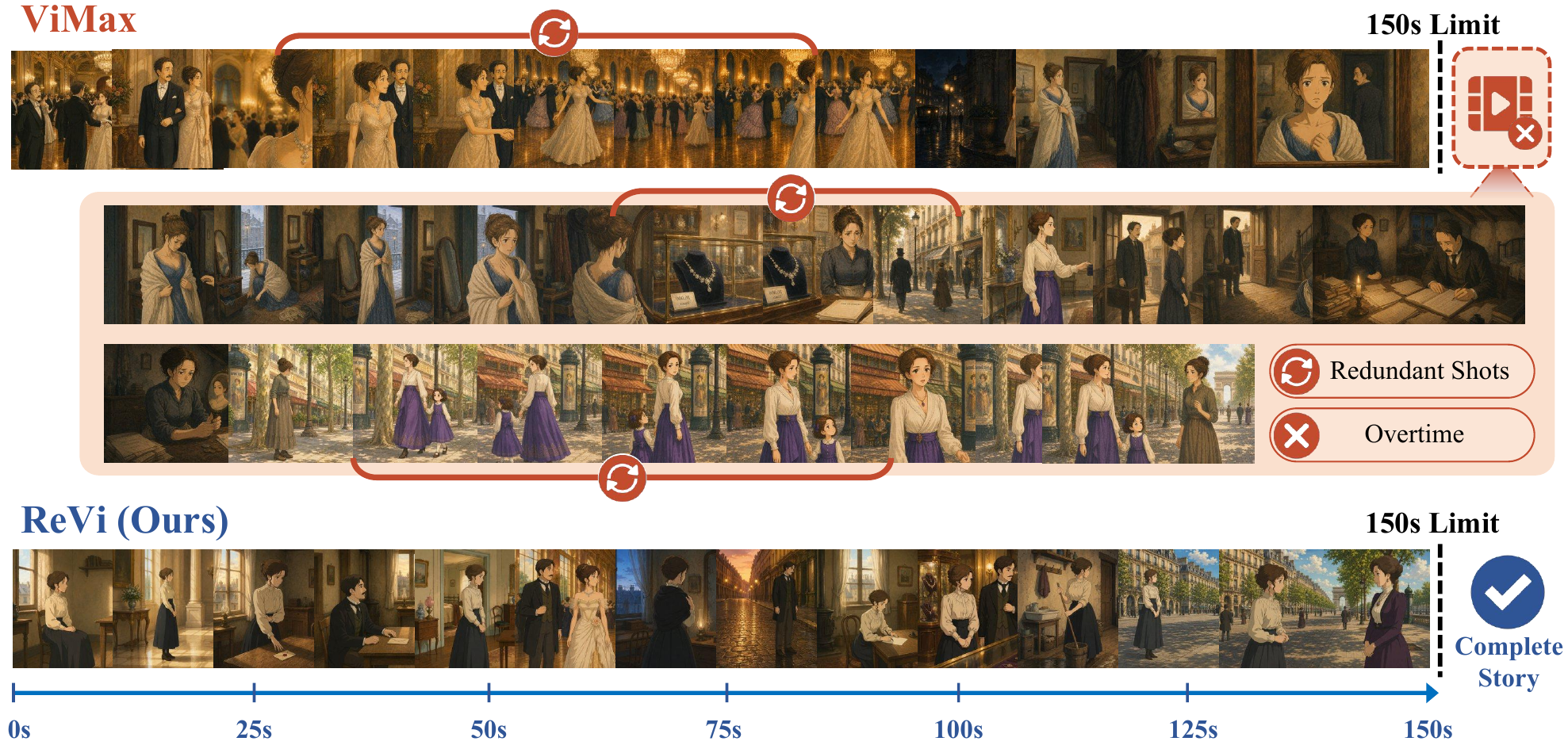}
\caption{\textbf{Story progression in \emph{The Necklace}.} ReVi completes
the story within the duration contract; ViMax exceeds it
and repeats shots marked by red outlines.}
\label{fig:case1}
\end{figure}

\subsection{Recursive Policy Improvement}
\label{sec:outer_loop_experiments}

Figure~\ref{fig:outer_loop}(a) shows that recap quality rises as the editing
policy is revised, while the Frozen condition remains unchanged. Allowing
meta-policy revision produces a steeper trajectory and a higher final quality
level. The gains on six held-out Recap32 novels are consistent with policy changes that
carry beyond the recaps used during search. The run-paired comparisons in Figure~\ref{fig:outer_loop}(b) show positive
effects from editing-policy revision across RF, NP, and AVG. Meta-policy revision adds gains in each dimension.
Figure~\ref{fig:outer_loop}(c) further shows greater quality improvement at
comparable candidate counts when the meta policy can evolve, indicating more
productive editor search under the measured protocol. Duration error also
declines across the conditions while evaluated outputs continue to satisfy
the duration contract. Together, these results support ReVi's bounded
recursive self-improvement claim at both the editing-policy and meta-policy
levels.

\begin{figure}[H]
\centering
\includegraphics[width=\linewidth]{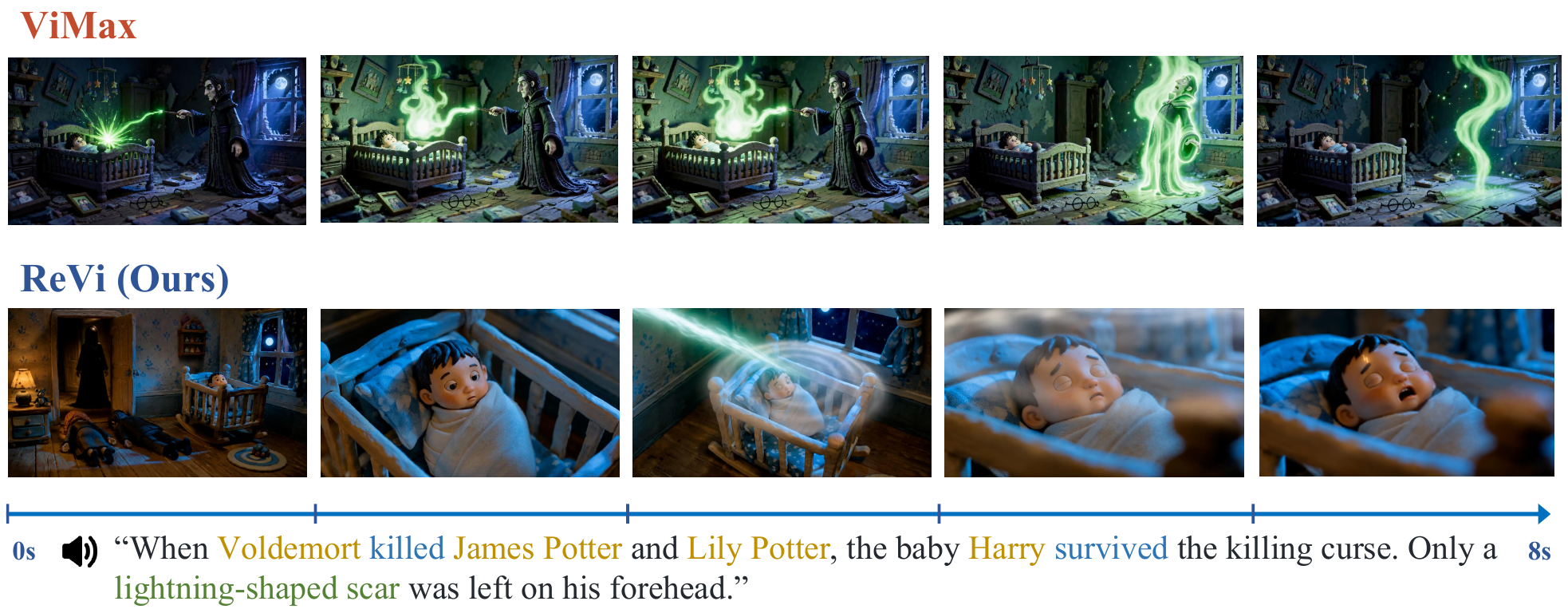}
\caption{\textbf{Comparison of shots in generated recaps.}
Frames partly selected from the generated recap video of \emph{Harry Potter and the Philosopher's Stone}.}
\label{fig:case2}
\end{figure}

\subsection{Qualitative Case Study}
\label{sec:case_study}
Figure~\ref{fig:case1} compares complete recaps of \emph{The Necklace}
against a 150-second target. ReVi reaches the concluding encounter within
the budget; ViMax repeats highlighted shots and runs over time. This
illustrates how event selection and timing affect story progression.

Figure~\ref{fig:case2} shows an eight-second passage from \emph{Harry Potter
and the Philosopher's Stone}. ViMax holds a wide view of Voldemort and the
crib, while ReVi cuts from the fallen adults and infant Harry to the curse
and Harry's reaction, matching distinct narrated Dramatic Beats.


\section{Conclusion}

We introduced ReVi for fixed-duration audiovisual recaps of novels. Its
inner loop plans causal coverage, timed narration, and matching shots;
its outer loop revises reusable editing and meta policies. Across two
evaluation sets, ReVi improves measured quality and duration compliance.
Human ratings and repeated policy-search results provide further evidence.
Future work should test more novels and independent generations, and
strengthen human assessment of recap fidelity.


\subsection*{AI use statement}

Generative AI models are components of the video-production pipelines and
automated evaluation described in Section~\ref{sec:experimental_setup}.
We use AI to polish paper writing and assist with finding related work. The authors take responsibility for the
methodology, reported results, figures, and final text.





\bibliography{iclr2027_conference}
\bibliographystyle{iclr2027_conference}


\appendix
\section{Related Work}

\paragraph{Narrative Video Generation}

Narrative video generation extends prompt-conditioned synthesis~
\citep{yang2025cogvideox,wan2025video} toward variable-length sequences~
\citep{villegas2023phenaki}, holistic multi-shot generation~
\citep{meng2026holocine}, and memory-conditioned shot synthesis~
\citep{zhang2025storymem}. Story visualization addresses visual consistency
across textual narratives~
\citep{li2019storygan,gong2023talecrafter,zhou2024storydiffusion}.
Language-guided planning~\citep{lin2024videodirectorgpt} and multi-agent
production~\citep{hu2024storyagent,wu2025movieagent,wang2026mavis,huang2026vimax}
connect these capabilities to coordinated storytelling workflows, with
specialized approaches to cinematic control~\citep{huang2026filmaster,liu2026realjailbreakfinegrainedjailbreak,liu2026clawkeepercomprehensivesafetyprotection,liu2025scalesjustitiacomprehensivesurvey},
world-state modeling~\citep{zuo2026filmworld}, and narrated storybooks~
\citep{xu2025mmstoryagent}. Feedback-driven systems further refine video
prompts~\citep{long2026vista} or combine global creative-direction search
with local multimodal refinement~\citep{song2026codirector,luo2026distilling,luo2026otk}. ReVi shares
these planning and feedback mechanisms, but focuses on compressing an
existing novel into a self-contained recap, jointly constraining causal
content, narration duration, and visual coverage.

\begin{table}[t]
    \caption{Paper Edit operations and the information they preserve.}
    \label{tab:paper_edit_operations}
    \centering
    \begingroup
    \footnotesize
    \setlength{\tabcolsep}{2pt}
    \setlength{\arrayrulewidth}{0.3pt}
    \renewcommand{\arraystretch}{1.10}
    \setlength{\aboverulesep}{1.5pt}
    \setlength{\belowrulesep}{1.5pt}
    \arrayrulecolor{black}
    \begin{tabular}{@{}
        >{\raggedright\arraybackslash}p{0.15\linewidth}|
        >{\raggedright\arraybackslash}p{0.37\linewidth}|
        >{\raggedright\arraybackslash}p{0.39\linewidth}@{}}
        \toprule[0.8pt]
        Operation & Editorial function & Information to preserve \\
        \specialrule{0.4pt}{1.5pt}{1.2pt}
        \textsc{Keep}
        & Retain an event as a distinct unit.
        & Its necessary action, motivation, or outcome. \\
        \textsc{Compress}
        & Shorten an event or a contiguous event chain.
        & Entry conditions, consequential changes, and outcomes. \\
        \textsc{Omit}
        & Remove material from the recap.
        & The prerequisites of subsequent retained events. \\
        \textsc{Merge}
        & Combine overlapping or repetitive material.
        & Distinct causal contributions of the combined events. \\
        \textsc{Bridge}
        & Insert a concise, source-supported connection across omitted material.
        & The relation needed to understand the transition. \\
        \textsc{Externalize}
        & Assign an internal state or implicit motivation to narration.
        & Its narrative role without inventing an observable action. \\
        \specialrule{0.4pt}{1.5pt}{1.2pt}
        \specialrule{0.8pt}{0pt}{0pt}
    \end{tabular}
    \endgroup
\end{table}

\paragraph{Narrative Summarization}

Narrative summarization selects essential content while accounting for
long-range causal and temporal dependencies. BOOKSUM studies summaries at
paragraph, chapter, and book levels~\citep{kryscinski2022booksum}, while
screenplay summarization uses narrative structure to identify salient
content~\citep{papalampidi2020screenplay}. PlotSnap and TaleSumm
use television recaps to study summarization of existing episodes~
\citep{singh2024previously}, and MovieRecapsQA evaluates open-ended question
answering from recap visuals and aligned movie dialogue~
\citep{shaar2026movierecapsqa}. ReVi extends the content-selection
to generating new narration and footage directly from a novel, requiring
explicit temporal allocation and audiovisual realization in a fixed runtime.

\paragraph{Recursive Self-Improvement.} 

Self-improvement ranges from revising individual outputs~
\citep{madaan2023selfrefine,long2026vista} to optimizing reusable prompts~
\citep{agrawal2026gepa,wang2026lancetneuralinterventionstructural} and agent designs~\citep{hu2025automated}. STOP and
the Darwin G\"odel Machine explore recursive program improvement and coding
agent evolution~\citep{zelikman2024stop,zhang2026darwin,chen2026securingcomputeruseagentsunified}. Promptbreeder
improves task prompts and their mutation prompts~
\citep{fernando2024promptbreeder}, while Hyperagents makes the task
agent and its improvement mechanism editable~\citep{zhang2026hyperagents}.
ReVi instantiates this distinction for recap editing: the Inner Loop repairs
the current video, and the Outer Loop revises editing and meta policies using
experience across novels. Given reported limitations of intrinsic
self-correction on reasoning tasks~\citep{huang2024selfcorrect}, a frozen
external Evaluation Kernel governs successor acceptance, and meta policies
are compared through the editors they produce under matched budgets.

\section{Additional Method Details}
\label{sec:method_details}

\subsection{Complete Algorithm}
\label{sec:full_algorithm}

\begin{algorithm}[H]
    \caption{ReVi: editing and recursive policy improvement}
    \label{alg:revi}
    \small
    \begin{algorithmic}[1]
        \Require Development/promotion data $\mathcal{D}_{\mathrm{dev}},\mathcal{D}_{\mathrm{prom}}$; initial policies $\Theta_0$;
        \Statex \hspace{\algorithmicindent} meta-policy episodes $\mathcal{E}_{\mathrm{meta}}$; frozen kernel $E^*$;
        \Statex \hspace{\algorithmicindent} production/evolution budgets and search budget $K$
        \Ensure Accepted policy state $\Theta_t=(\theta_t,\mu_t)$
        \State $t\gets 0$; $\Theta_t=(\theta_t,\mu_t)\gets\Theta_0$; trace buffer $\mathcal{T}\gets\emptyset$; candidate archive $\mathcal{A}\gets\emptyset$
        \While{evolution budget remains}
            \ForAll{$(x,c)\in\mathcal{D}_{\mathrm{dev}}$} \Comment{Inner Loop: fixed $\theta_t$}
                \State $G \gets \Call{StoryMapper}{x,c;\theta_t}$ \Comment{source events $e_i$}
                \State $P \gets \Call{PaperEditor}{G,c;\theta_t}$ \Comment{edit units $p_i$}
                \State $Z \gets \Call{RadioEditor}{P,c;\theta_t}$ \Comment{timed segments $z_i$}
                \State $Q \gets \Call{AVPlanner}{G,Z,c;\theta_t}$ \Comment{beats $b_i$, requests $q_j$}
                \State $(y,F) \gets \Call{RenderCheck}{G,P,Z,Q,c}$
                \While{$F\neq\emptyset$ and repair budget remains}
                    \State Revise the earliest responsible stage under $\rho_t$; recompute dependent artifacts
                    \State $(y,F) \gets \Call{RenderCheck}{G,P,Z,Q,c}$
                \EndWhile
                \State Record the production trace in $\mathcal{T}$, including unsuccessful attempts
            \EndFor
            \State $\theta_{t+1} \gets \operatorname{Search}_{\mu_t}(\theta_t,\mathcal{T},\mathcal{A};K)$
            \Statex \hspace{\algorithmicindent}\Comment{test $\widetilde{\theta}_t^{(i)}$ with $E^*$ on $\mathcal{D}_{\mathrm{prom}}$; retain $\theta_t$ if none passes}
            \State $\mu'_t \gets \Call{ProposeMeta}{\mu_t,\mathcal{T},\mathcal{A}}$ \Comment{a $\widetilde{\mu}_t^{(j)}$ candidate}
            \State Evaluate $V(\mu'_t)$ and $V(\mu_t)$ on $\mathcal{E}_{\mathrm{meta}}$ with matched inputs and budget $K$
            \State $\mu_{t+1} \gets \operatorname{Promote}_{E^*}(\mu_t,\mu'_t;V)$
            \State Archive tried candidate states $\widetilde{\Theta}_t^{(i)}$ and their outcomes in $\mathcal{A}$
            \State $\Theta_{t+1}\gets(\theta_{t+1},\mu_{t+1})$; $t\gets t+1$
        \EndWhile
        \State \Return $\Theta_t$
    \end{algorithmic}
\end{algorithm}

\subsection{Editorial Operations and Narration Modes}

Table~\ref{tab:paper_edit_operations} defines the transformations recorded
in the Paper Edit. Removing presentation detail is permitted when the
information needed to understand retained developments survives elsewhere
in the account. In particular, \textsc{Bridge} supplies a source-supported
connection across omitted material, while \textsc{Externalize} assigns an
internal state or implicit motivation to narration without inventing an
observable action. The closure checker examines prerequisites,
setup--payoff relations, source support, and disclosure order after revision.

Table~\ref{tab:radio_edit_operations} gives the operational adaptations of
Genette's duration categories. These modes determine the granularity of
narration under the recap budget. The \textsc{Scene} adaptation permits
comparatively detailed treatment without requiring exact equality between
story time and playback time. The Radio Editor may elaborate retained
material when time remains, but cannot independently restore events
excluded by the Paper Edit.

\begin{table}[t]
    \caption{Narration modes adapted from Genette's account of narrative duration.}
    \label{tab:radio_edit_operations}
    \centering
    \begingroup
    \footnotesize
    \setlength{\tabcolsep}{2pt}
    \setlength{\arrayrulewidth}{0.3pt}
    \renewcommand{\arraystretch}{1.10}
    \setlength{\aboverulesep}{1.5pt}
    \setlength{\belowrulesep}{1.5pt}
    \arrayrulecolor{black}
    \begin{tabular}{@{}
        >{\raggedright\arraybackslash}p{0.16\linewidth}|
        >{\raggedright\arraybackslash}p{0.80\linewidth}@{}}
        \toprule[0.8pt]
        Mode & Role in recap narration \\
        \specialrule{0.4pt}{1.5pt}{1.2pt}
        \textsc{Ellipsis}
        & Pass over an interval already excluded by the Paper Edit, using a
          brief temporal connection when needed. \\
        \textsc{Summary}
        & Condense a retained event group into economical narration. \\
        \textsc{Scene}
        & Give a key event a more detailed, comparatively direct presentation. \\
        \textsc{Pause}
        & Explain a necessary motivation, rule, or background fact without
          advancing the narrated action. \\
        \specialrule{0.4pt}{1.5pt}{1.2pt}
        \specialrule{0.8pt}{0pt}{0pt}
    \end{tabular}
    \endgroup
\end{table}

\subsection{Policy Revisions and Search Accounting}

Table~\ref{tab:editing_policy_targets} summarizes the persistent targets
of editing-policy revision. A candidate patch records its target component,
triggering condition, changed instruction or configuration, expected effect,
and constraints. Single-component changes are preferred when they isolate
the suspected cause; coordinated changes to adjacent stages are considered
when evidence implicates their interaction.

\begin{table}[t]
    \caption{Persistent modification targets of the editing policy.}
    \label{tab:editing_policy_targets}
    \centering
    \begingroup
    \footnotesize
    \setlength{\tabcolsep}{2pt}
    \setlength{\arrayrulewidth}{0.3pt}
    \renewcommand{\arraystretch}{1.10}
    \setlength{\aboverulesep}{1.5pt}
    \setlength{\belowrulesep}{1.5pt}
    \arrayrulecolor{black}
    \begin{tabular}{@{}
        >{\raggedright\arraybackslash}p{0.27\linewidth}|
        >{\raggedright\arraybackslash}p{0.69\linewidth}@{}}
        \toprule[0.8pt]
        Policy component & Persistent modification target \\
        \specialrule{0.4pt}{1.5pt}{1.2pt}
        Source interpretation and causal compression
        & Rules for identifying necessary motivations and dependencies,
          prioritizing retained events, and choosing \textsc{Keep},
          \textsc{Compress}, \textsc{Omit}, \textsc{Merge}, \textsc{Bridge},
          or \textsc{Externalize}. \\
        Temporal allocation
        & Rules for assigning duration modes, balancing information density,
          distributing telling time, and requesting structural compression. \\
        Audiovisual coordination
        & Rules for dividing information between narration and visuals,
          defining Beat boundaries, selecting visual coverage, and grouping
          Beats into feasible requests. \\
        Repair
        & Conditions for local correction, upstream revision, regeneration,
          and reuse of unaffected artifacts. \\
        \specialrule{0.4pt}{1.5pt}{1.2pt}
        \specialrule{0.8pt}{0pt}{0pt}
    \end{tabular}
    \endgroup
\end{table}

Structural and narration checks can reject candidates with broken
dependencies or infeasible speech duration before rendering. These checks
allocate evaluation resources; they do not replace assessment of the final
video. The frozen Evaluation Kernel applies fixed acceptance constraints,
a quality--cost selection rule, and a promotion margin. A search that finds
no qualifying candidate retains its parent policy.

Production traces retain inputs, decisions, artifacts, critic feedback,
repairs, and outcomes. The improvement archive separately records the
parent policy, diagnosed cause, patch, expected benefit, evaluation outcome,
regressions, and cost of each attempt. The Meta-Agent can inspect these
records but cannot rewrite them. This separation allows meta-policy
revision to use evidence about both production failures and unsuccessful
attempts to correct them.

For meta-policy comparisons, each candidate receives the same parent
editor, accessible traces, improvement history, and total search budget.
Search accounting includes diagnosis, proposal generation, candidate
evaluation, and rendering. Meta-policy development and comparison costs
are recorded separately. Editor selection uses the episode's development
and promotion data; the selected editor is frozen before evaluation on its
held-out novels. The protocol reserves a sealed test partition outside
policy search for assessing the resulting system. Foundation models,
source novels, task contracts, evaluation definitions, and final acceptance
rules remain outside the mutable policies throughout evolution.

\section{Additional Experimental Details}

\label{sec:experimental_provenance}

\paragraph{Datasets.}
We use FilmEval's 15 narrative texts and its Easy, Medium, and Hard
difficulty tiers~\citep{zuo2026filmworld}, with five sources per tier.
For the recap-generation protocol, we assign duration targets of 90,
120, and 150 seconds to the respective tiers.
For Recap32, we selected 32 book-length English
Gutenberg entries from NarrativeQA~\citep{kocisky2018narrativeqa}.
The 12/4/4/12 allocation specifies evolution, editor promotion,
meta-policy evaluation, and testing, with a 180-second target.
Table~\ref{tab:recap32_candidates} lists the selected titles and maps
test IDs N01--N12 to novels in the listed order.
The reported main comparison covers these 12 test IDs; the other
20 titles are assigned to policy-development partitions.

\begin{table}[t]
\centering
\caption{Recap32 novel allocation. Test IDs N01--N12 follow the listed title order; main-comparison results cover these 12 novels.}
\label{tab:recap32_candidates}
\footnotesize
\begin{tabular}{@{}p{0.20\linewidth}p{0.75\linewidth}@{}}
\toprule
Partition & Titles \\
\midrule
Evolution (12) & \emph{Percival Keene}; \emph{The Rise of Silas Lapham}; \emph{Lorna Doone}; \emph{Five Children and It}; \emph{Amelia}; \emph{Jennie Gerhardt}; \emph{Anne of Green Gables}; \emph{The Prairie}; \emph{The Odd Women}; \emph{Moonfleet}; \emph{The Riddle of the Sands}; \emph{The Fifth Queen}. \\
Editor promotion (4) & \emph{Brewster's Millions}; \emph{Beyond the Rocks}; \emph{The Mysterious Affair at Styles}; \emph{Rebecca of Sunnybrook Farm}. \\
Meta evaluation (4) & \emph{The Golden Bowl}; \emph{The Iron Heel}; \emph{The Mysterious Island}; \emph{The Mill on the Floss}. \\
Test (12) & \texttt{N01} \emph{Marianela}; \texttt{N02} \emph{The Hour of the Dragon}; \texttt{N03} \emph{Lothair}; \texttt{N04} \emph{Indiscretions of Archie}; \texttt{N05} \emph{Seventeen}; \texttt{N06} \emph{McTeague}; \texttt{N07} \emph{Comrades: A Story of Social Adventure in California}; \texttt{N08} \emph{Moll Flanders}; \texttt{N09} \emph{The Fortune of the Rougons}; \texttt{N10} \emph{The House of the Seven Gables}; \texttt{N11} \emph{Tarzan of the Apes}; \texttt{N12} \emph{Mary Barton}. \\
\bottomrule
\end{tabular}
\end{table}

\paragraph{Metrics.}
All three quality scores lie in $[0,100]$.
\textbf{Recap Fidelity (RF)} is defined as the mean of weighted
essential-event coverage (WEC), logic reliability (LR), and hallucination
resistance (NHR), using a shared duration-specific source reference.
WEC gives full, half, or zero credit to covered, partially covered, or
missing events, with event weights of 1, 2, or 3.
LR and NHR assess logical errors and unsupported content, respectively.
\textbf{Narrative Pacing (NP)} evaluates narrative clarity, progression,
transitions, and pacing. It accepts fine-grained scores from 0 to 100;
the five rubric anchors are not a five-value output scale.
\textbf{Audio-Visual Grounding (AVG)} is the percentage of narration
segments labeled \emph{Support} or \emph{Complement}, relative to all
expected segments. It is not the average of the three quality metrics.
For final duration $L_i$ and target $T_i$, \textbf{Duration Error (DE)} is
$100|L_i-T_i|/T_i$. In the reported data, \textbf{Duration Pass Rate (DP)} is the
percentage of all records with a complete output and DE at most 5\%.
All reported outputs completed.

\paragraph{Baselines and implementation.}
ViMax~\citep{huang2026vimax} is the primary system-level baseline.
The FilmEval comparison additionally includes
VideoClaw~\citep{hitsztmg2026videoclaw},
MovieAgent~\citep{wu2025movieagent},
MM-StoryAgent~\citep{xu2025mmstoryagent}, and
VGoT~\citep{zheng2025videogenofthought}.
\emph{Direct LLM} generates narration and a storyboard in one pass before
using the common rendering pipeline. Recap32 compares Direct LLM, ViMax,
MovieAgent, and ReVi. Baselines receive the requested duration in their
prompts, but their original planning and rendering pipelines do not enforce
a duration constraint. The experimental configuration uses
\texttt{gpt-5.6-terra} for planning, \texttt{gpt-image-2} for images,
Seedance~2.0 (\texttt{doubao-seedance-2-0-260128}) for video, and Mureka TTS
(\texttt{mureka/tts}, Emma, default rate) for narration. The shared output
specification is 720p, 16:9, and a Ghibli-style prompt, with video requests
of 1--15 seconds. ReVi permits two regenerations per asset, first frame,
or video segment, and three cross-stage repairs per source. Adapters
supply plot briefs and narration where required by a baseline.

\paragraph{Scoring and uncertainty.}
We use three Gemini~3.7 Flash evaluations of each final output. These
calls are repeated judgments, not independent generations. Individual
RF component judgments are unavailable for recomputation.
Means weight sources equally and are rounded only for display.
The source-paired analyses below quantify variation across source
texts, not repeated-generation uncertainty.

\paragraph{Inner-loop interventions.}
We ablate five mechanisms on C1, C3, C6, C8, C11, and C13, two sources
per difficulty tier. Each variant changes one mechanism at the same
target duration. Each variant is compared with the six Full ReVi
records from the same ablation batch. \emph{Causal Editing} replaces causal-chain
selection and dependency repair with ordinary summary editing;
\emph{TTS Feedback} replaces measured speech duration with word-count
estimates; \emph{Narrative Allocation} replaces function-aware modes
and time allocation with a single length budget while retaining TTS
feedback; \emph{Beat Planning} groups visuals by sentence boundaries
instead of event changes; and \emph{Repair Routing} confines revisions
to the stage reporting the error.
\paragraph{Record linkage and source metadata.}
The main comparison comprises 105 FilmEval and 48 Recap32 video-level
observations. The inner-loop analysis comprises 36 observations, and
the repeated-search analysis comprises 1,080 observations across
conditions, rounds, runs, and sources. These are distinct experimental
batches: the six Full ReVi scores in the ablation batch differ from
those for the same FilmEval IDs in the main comparison. We therefore
make ablation contrasts within that batch and analyze repeated-search
initial scores separately from the main comparison.

The title mapping links N01--N12 to the test novels in
Table~\ref{tab:recap32_candidates}.

\paragraph{RF aggregation and segment scoring.}
Each Gemini evaluation scores RF from weighted essential-event coverage, logic
reliability, and hallucination resistance. The reported video-level RF
is the median of the three complete RF judgments.
NP uses a fine-grained 0--100 scale, with 0/25/50/75/100 as qualitative
anchors. AVG counts \emph{Support} and \emph{Complement} segments as
positive; its denominator is all expected narration segments. Segment
labels and per-call judgments are unavailable for independent rescoring.

\paragraph{Resampling and agreement.}
The source ID is the resampling unit, and each replicate retains all
paired methods or rounds. FilmEval resamples five sources within
each difficulty tier; inner-loop and human comparisons resample
two per tier. Recap32 main comparisons resample its 12 source IDs
without strata. Outer-loop intervals resample independent search runs
while preserving paired condition outcomes.
Each analysis uses 10,000 replicates, NumPy's default generator with
seed 20260921, and linear-interpolated 2.5th/97.5th percentiles.
Main-comparison intervals are pointwise and quantify source-sampling
uncertainty. Outer-loop intervals quantify variation across search runs
on the six-source subset.
The human analysis keeps all seven methods together within each of
six source clusters. Spearman correlations recompute average ranks
within each replicate. Regression bands refit human score on automatic
score in each replicate and describe the fitted mean, not individual
predictions. No independent-video $p$-values are reported.

Ordinal Krippendorff $\alpha$ uses the ordered domain $\{0,1,2,3,4,5\}$
and all three ratings on each of the 42 videos. For pooled category
counts $n_c$, the squared ordinal distance between categories $a<b$ is
$[\sum_{c=a}^{b}n_c-(n_a+n_b)/2]^2$. Observed within-video pairs
are weighted by $1/(R_i-1)$, and expected pairs use pooled category
counts. We compute $\alpha=1-D_o/D_e$. No ratings are missing.
Rating SD uses the sample denominator 17 for each method's 18 ratings;
it is distinct from the SD of the six video means.

\paragraph{Cost accounting and rounding.}
Main-comparison expenditure uses per-video generation costs.
Repeated-search results report evaluated candidate entries as the measure
of search volume in Figure~\ref{fig:outer_loop}(c).
Quality means and paired differences are calculated from per-video
values before display rounding. Dollar values use decimal half-up
rounding, and pass rates retain integer counts.
DE uses the reported higher-precision field; runtime rounding to
0.01 seconds accounts for its small recomputation discrepancy.

\section{Additional Results}
\label{sec:additional_experimental_results}

\subsection{Shared-Narration Duration Control}
\label{sec:shared_narration_control}
For each source, a human editor prepared one narration script at the
requested duration. All methods used the same audio file for that source
to generate their recap videos. This control fixes spoken content and
audio timing while retaining each method's video selection, alignment,
and rendering. Table~\ref{tab:equal_duration} reports the completed
videos on the same FilmEval and Recap32 sources as the main comparison.
Final playback length can still vary across methods, so we report both
its mean and the duration-pass count. ReVi retains higher RF, NP, and
AVG scores than ViMax under shared narration: the source-paired gains
are 6.47/6.12/6.18 points on FilmEval and 6.02/6.72/6.72 points on
Recap32, respectively. This control isolates downstream recap-video
realization under fixed narration; the main comparison evaluates the
complete editing pipeline. As a sensitivity check, restricting to pairs
where both methods pass the 5\% duration tolerance leaves 13 FilmEval
and 11 Recap32 sources; ReVi's RF/NP/AVG gains over ViMax remain
6.63/6.00/6.46 and 6.19/6.68/6.86 points, respectively.

\begin{table}[H]
\centering
\caption{Quality under a shared narration track.}
\label{tab:equal_duration}
\begingroup
\footnotesize
\setlength{\tabcolsep}{0pt}
\setlength{\arrayrulewidth}{0.3pt}
\renewcommand{\arraystretch}{1.10}
\setlength{\aboverulesep}{1.5pt}
\setlength{\belowrulesep}{1.5pt}
\arrayrulecolor{black}
\begin{tabular}{@{}>{\raggedright\arraybackslash\hspace{2pt}}p{0.220\linewidth}|>{\centering\arraybackslash}p{0.105\linewidth}>{\centering\arraybackslash}p{0.105\linewidth}>{\centering\arraybackslash}p{0.105\linewidth}|>{\centering\arraybackslash}p{0.105\linewidth}|>{\centering\arraybackslash}p{0.110\linewidth}>{\centering\arraybackslash}p{0.240\linewidth}@{}}
\toprule[0.8pt]
Method & RF $\uparrow$ & NP $\uparrow$ & AVG $\uparrow$ & $L$ (s) & DE $\downarrow$ & DP $\uparrow$ \\
\specialrule{0.4pt}{1.5pt}{1.2pt}
\multicolumn{7}{c}{\textit{FilmEval: 15 sources; 90/120/150 s}} \\
\midrule[0.4pt]
Direct LLM & 72.97 & 69.61 & 67.80 & 117.8 & 4.16 & 80.0\% (12/15) \\
ViMax & \underline{84.33} & \underline{82.34} & \underline{81.11} & 121.3 & \underline{3.12} & \underline{86.7\% (13/15)} \\
VideoClaw & 80.98 & 79.02 & 78.29 & 120.5 & 4.15 & 73.3\% (11/15) \\
MovieAgent & 82.56 & 79.66 & 76.49 & 118.8 & 4.23 & 66.7\% (10/15) \\
MM-StoryAgent & 76.46 & 74.43 & 72.04 & 118.1 & 4.06 & 66.7\% (10/15) \\
VGoT & 78.25 & 76.58 & 74.64 & 120.5 & 4.69 & 46.7\% (7/15) \\
\rowcolor{MainTableOurs}
\textbf{ReVi} & \textbf{90.81} & \textbf{88.47} & \textbf{87.30} & 119.9 & \textbf{1.95} & \textbf{100.0\% (15/15)} \\
\specialrule{0.4pt}{1.5pt}{1.2pt}
\specialrule{0.4pt}{0pt}{1.5pt}
\multicolumn{7}{c}{\textit{Recap32: 12 test novels; 180 s}} \\
\midrule[0.4pt]
Direct LLM & 72.38 & 69.96 & 67.51 & 178.9 & 4.54 & 41.7\% (5/12) \\
ViMax & \underline{83.41} & \underline{81.74} & \underline{80.10} & 178.6 & \underline{3.15} & \underline{91.7\% (11/12)} \\
MovieAgent & 82.18 & 79.34 & 76.44 & 180.7 & 3.43 & 83.3\% (10/12) \\
\rowcolor{MainTableOurs}
\textbf{ReVi} & \textbf{89.43} & \textbf{88.46} & \textbf{86.82} & 180.6 & \textbf{1.60} & \textbf{100.0\% (12/12)} \\
\specialrule{0.4pt}{1.5pt}{1.2pt}
\specialrule{0.8pt}{0pt}{0pt}
\end{tabular}
\par\vspace{4pt}
\begin{minipage}{\linewidth}
\footnotesize
For each source, every method uses the same human-edited narration script and identical audio file. RF/NP/AVG are quality scores; $L$ is mean final playback length; DE is mean absolute duration error in percent; DP is the share within 5\% of the target (pass counts in parentheses). Rankings are within each dataset. Bold and underline mark the best and the second-best results, respectively; ties share a rank.
\end{minipage}
\endgroup
\end{table}

\subsection{Inner Loop Ablations}
\label{sec:inner_loop_ablations}
We ablate five mechanisms on six FilmEval texts, two per tier,
at the main-comparison target durations. The replacements use ordinary
summaries for Causal Editing, word-count timing for TTS Feedback, one
narration budget for Narrative Allocation, sentence boundaries for Beat
Planning, and stage-local revisions for Repair Routing.
Table~\ref{tab:inner_loop_ablations} compares each variant with the
Full ReVi records from the same ablation batch. These scores are kept
separate from the main-comparison ReVi scores.
\begin{table}[t]
\centering
\caption{Inner-loop component ablations.}
\label{tab:inner_loop_ablations}
\begingroup
\footnotesize
\setlength{\tabcolsep}{0pt}
\setlength{\arrayrulewidth}{0.3pt}
\renewcommand{\arraystretch}{1.10}
\setlength{\aboverulesep}{1.5pt}
\setlength{\belowrulesep}{1.5pt}
\arrayrulecolor{black}
\begin{tabular}{@{}>{\raggedright\arraybackslash\hspace{2pt}}p{0.230\linewidth}|>{\centering\arraybackslash}p{0.140\linewidth}>{\centering\arraybackslash}p{0.140\linewidth}>{\centering\arraybackslash}p{0.140\linewidth}|>{\centering\arraybackslash}p{0.130\linewidth}>{\centering\arraybackslash}p{0.210\linewidth}@{}}
\toprule[0.8pt]
Ablation & RF $\uparrow$ & NP $\uparrow$ & AVG $\uparrow$ & DE $\downarrow$ & DP $\uparrow$ \\
\specialrule{0.4pt}{1.5pt}{1.2pt}
\multicolumn{6}{c}{\textit{Paper Editor}} \\
\midrule[0.4pt]
w/o Causal Editing & 87.10 & \underline{88.92} & \underline{88.39} & \underline{2.18} & \textbf{100.0\% (6/6)} \\
\specialrule{0.4pt}{1.5pt}{1.2pt}
\specialrule{0.4pt}{0pt}{1.5pt}
\multicolumn{6}{c}{\textit{Radio Editor}} \\
\midrule[0.4pt]
w/o TTS Feedback & \underline{91.68} & 87.71 & 86.82 & 8.65 & 0.0\% (0/6) \\
w/o Narrative Allocation & 89.84 & 85.17 & 86.27 & 3.69 & \underline{83.3\% (5/6)} \\
\specialrule{0.4pt}{1.5pt}{1.2pt}
\specialrule{0.4pt}{0pt}{1.5pt}
\multicolumn{6}{c}{\textit{AV Planner}} \\
\midrule[0.4pt]
w/o Beat Planning & 90.79 & 88.23 & 79.99 & 2.21 & \textbf{100.0\% (6/6)} \\
\specialrule{0.4pt}{1.5pt}{1.2pt}
\specialrule{0.4pt}{0pt}{1.5pt}
\multicolumn{6}{c}{\textit{Cross-agent orchestration}} \\
\midrule[0.4pt]
w/o Repair Routing & 87.23 & 87.42 & 85.75 & 3.44 & \underline{83.3\% (5/6)} \\
\specialrule{0.4pt}{1.5pt}{1.2pt}
\specialrule{0.4pt}{0pt}{1.5pt}
\multicolumn{6}{c}{\textit{Full system}} \\
\midrule[0.4pt]
\rowcolor{MainTableOurs}
\textbf{Full ReVi} & \textbf{91.87} & \textbf{90.52} & \textbf{88.92} & \textbf{1.86} & \textbf{100.0\% (6/6)} \\
\specialrule{0.4pt}{1.5pt}{1.2pt}
\specialrule{0.8pt}{0pt}{0pt}
\end{tabular}
\par\vspace{4pt}
\begin{minipage}{\linewidth}
\footnotesize
Inner-loop ablations on six sources, grouped by the editing agent that owns each local mechanism. Repair Routing evaluates cross-agent orchestration. Bold and underline mark the best and the second-best results, respectively; ties share a rank.
\end{minipage}
\endgroup
\end{table}

\paragraph{Causal selection and narration allocation.}
Removing Causal Editing produces the largest RF loss, 4.76 points, while
NP falls by 1.60 points. This concentration is consistent with the Paper
Editor's role of retaining causally necessary events rather than merely
shortening the plot. Removing Narrative Allocation most affects NP
($-5.35$ points), consistent with the need to reserve narration time
according to each retained event's function. These interventions affect
different aspects of the story plan: deciding what information survives
compression and deciding how long the retelling spends on it.

\paragraph{Visual beats and measured speech timing.}
Replacing event beats with sentence boundaries yields the largest AVG
loss ($-8.93$ points). The spoken sentences remain available, but the
visual plan no longer follows event changes as closely. Measured speech
timing addresses a different failure mode: without TTS Feedback, RF
changes by only $-0.18$ points, but DE rises by 6.79 percentage points
and duration compliance falls from 6/6 to 0/6. Plausible content
selection alone therefore does not ensure that the retelling fits the
viewing-time contract.

\paragraph{Cross-stage repair.}
Removing Repair Routing lowers RF and AVG by 4.63 and 3.17 points and
leaves one output outside the duration tolerance. This pattern is
consistent with errors requiring revisions across stages rather than
only at the stage that reports them. The interventions replace one
mechanism at a time on a limited source subset and do not identify how
pairs of mechanisms interact. Source-paired bounds are reported in
Table~\ref{tab:mechanism_intervals}.

\subsection{Repeated Outer-Loop Search}
\label{sec:rsi_details}
Table~\ref{tab:outer_results} reports condition--round
means for six Recap32 test novels (N02, N04, N06, N08, N10, and N12).
The analysis contains ten runs
per condition and six sources per run; every reported video completed
and passed the duration threshold. Each run reuses its Frozen output
for a source across rounds. The search log and policy timeline link each
accepted meta-policy version to one candidate that passed checks and was
promoted. Editing-Policy raises all three quality dimensions, and
Editing-Policy+Meta-Policy adds a further gain in the recorded
searches. Table~\ref{tab:outer_paired} reports final run-paired
contrasts. Editing-Policy minus Frozen is 2.20 [1.97, 2.41] RF points,
2.50 [2.30, 2.70] NP points, and 2.10 [1.89, 2.29] AVG points.
Editing-Policy+Meta-Policy minus Editing-Policy is 2.22 [2.04, 2.35],
1.99 [1.78, 2.19], and 2.41 [2.17, 2.63] points, respectively.
The intervals resample search runs, keeping the paired sources and
conditions together; they do not quantify uncertainty over a broader
population of novels. Duration pass rate is 100\% in all three
conditions, so the quality gains do not come from relaxing the stated
duration threshold. For the search-yield view in
Figure~\ref{fig:outer_loop}(c), we count every candidate entry in the
search records, including editor and meta-policy candidates.
Editing-Policy evaluates 147 candidates across ten searches, compared
with 195 for Editing-Policy+Meta-Policy. At the final checkpoint, their
mean gains in $Q=(\mathrm{RF}+\mathrm{NP}+\mathrm{AVG})/3$ over the
paired Frozen outputs are 2.26 and 4.47 points, respectively.
Figure~\ref{fig:outer_loop}(c) relates quality gains at each checkpoint
to the mean cumulative number of candidates evaluated per run.
\begin{table}[t]
\centering
\caption{Quality across recursive policy revisions.}
\label{tab:outer_results}
\begingroup
\footnotesize
\setlength{\tabcolsep}{0pt}
\setlength{\arrayrulewidth}{0.3pt}
\renewcommand{\arraystretch}{1.10}
\setlength{\aboverulesep}{1.5pt}
\setlength{\belowrulesep}{1.5pt}
\arrayrulecolor{black}
\begin{tabular}{@{}>{\raggedright\arraybackslash\hspace{2pt}}p{0.180\linewidth}|>{\centering\arraybackslash}p{0.195\linewidth}>{\centering\arraybackslash}p{0.195\linewidth}>{\centering\arraybackslash}p{0.195\linewidth}|>{\centering\arraybackslash}p{0.215\linewidth}@{}}
\toprule[0.8pt]
Round & RF $\uparrow$ & NP $\uparrow$ & AVG $\uparrow$ & DE $\downarrow$ \\
\specialrule{0.4pt}{1.5pt}{1.2pt}
\multicolumn{5}{c}{\textit{Frozen}} \\
\midrule[0.4pt]
0 & 91.05 & 88.56 & 82.87 & 2.35 \\
1 & 91.05 & 88.56 & 82.87 & 2.35 \\
2 & 91.05 & 88.56 & 82.87 & 2.35 \\
3 & 91.05 & 88.56 & 82.87 & 2.35 \\
4 & 91.05 & 88.56 & 82.87 & 2.35 \\
5 & 91.05 & 88.56 & 82.87 & 2.35 \\
\specialrule{0.4pt}{1.5pt}{1.2pt}
\specialrule{0.4pt}{0pt}{1.5pt}
\multicolumn{5}{c}{\textit{Editing policy}} \\
\midrule[0.4pt]
0 & 91.05 & 88.56 & 82.87 & 2.35 \\
1 & 91.82 & 89.42 & 83.58 & 2.23 \\
2 & 92.39 & 90.12 & 84.16 & 2.11 \\
3 & 92.76 & 90.54 & 84.48 & 2.06 \\
4 & 93.05 & 90.86 & 84.78 & 2.00 \\
5 & 93.24 & 91.05 & 84.97 & 1.95 \\
\specialrule{0.4pt}{1.5pt}{1.2pt}
\specialrule{0.4pt}{0pt}{1.5pt}
\multicolumn{5}{c}{\textit{Editing + meta policy}} \\
\midrule[0.4pt]
0 & 91.05 & 88.56 & 82.87 & 2.35 \\
1 & 92.60 & 90.10 & 84.43 & 2.13 \\
2 & 93.75 & 91.29 & 85.55 & 2.01 \\
3 & 94.52 & 92.08 & 86.41 & 1.92 \\
4 & 95.10 & 92.67 & 86.95 & 1.82 \\
\rowcolor{MainTableOurs}
5 & 95.46 & 93.05 & 87.38 & 1.76 \\
\specialrule{0.4pt}{1.5pt}{1.2pt}
\specialrule{0.8pt}{0pt}{0pt}
\end{tabular}
\par\vspace{4pt}
\begin{minipage}{\linewidth}
\footnotesize
Repeated-search evaluation on six Recap32 test novels. Values are pooled means across runs and sources; DE is duration error in percent.
\end{minipage}
\endgroup
\end{table}
\begin{table}[t]
\centering
\caption{Final paired quality differences.}
\label{tab:outer_paired}
\begingroup
\footnotesize
\setlength{\tabcolsep}{0pt}
\setlength{\arrayrulewidth}{0.3pt}
\renewcommand{\arraystretch}{1.10}
\setlength{\aboverulesep}{1.5pt}
\setlength{\belowrulesep}{1.5pt}
\arrayrulecolor{black}
\begin{tabular}{@{}>{\raggedright\arraybackslash\hspace{2pt}}p{0.300\linewidth}|>{\centering\arraybackslash}p{0.225\linewidth}>{\centering\arraybackslash}p{0.225\linewidth}>{\centering\arraybackslash}p{0.225\linewidth}@{}}
\toprule[0.8pt]
Contrast & RF & NP & AVG \\
\specialrule{0.4pt}{1.5pt}{1.2pt}
Editor -- Frozen & \shortstack[r]{+2.20\\{[+1.97, +2.41]}} & \shortstack[r]{+2.50\\{[+2.30, +2.70]}} & \shortstack[r]{+2.10\\{[+1.89, +2.29]}} \\
Editor + Meta -- Editor & \shortstack[r]{+2.22\\{[+2.04, +2.35]}} & \shortstack[r]{+1.99\\{[+1.78, +2.19]}} & \shortstack[r]{+2.41\\{[+2.17, +2.63]}} \\
\specialrule{0.4pt}{1.5pt}{1.2pt}
\specialrule{0.8pt}{0pt}{0pt}
\end{tabular}
\par\vspace{4pt}
\begin{minipage}{\linewidth}
\footnotesize
Final-round run-paired quality differences with run-bootstrap intervals. Each run contains the same six Recap32 source IDs under the compared conditions.
\end{minipage}
\endgroup
\end{table}

\subsection{Generation Cost}
\label{sec:generation_cost}
\begin{figure}[!htbp]
\centering
\includegraphics[width=\linewidth]{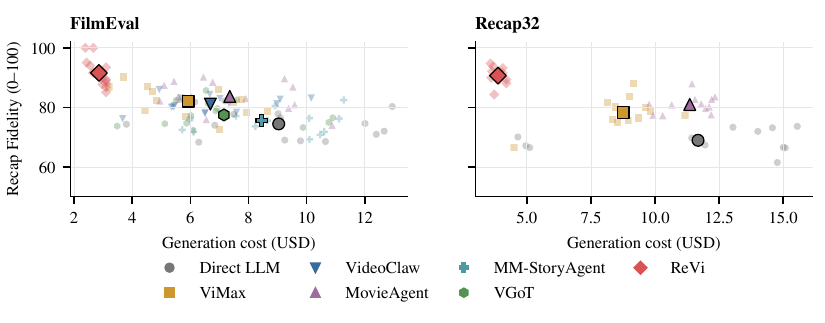}
\caption{\textbf{ReVi has higher mean Recap Fidelity (RF) at lower generation
cost than ViMax.} The comparison covers FilmEval and Recap32. Faint markers
show individual outputs and outlined markers show method means
(15 FilmEval and 12 Recap32 sources per method). Costs exclude
offline RSI search.}
\label{fig:cost_quality}
\end{figure}

Table~\ref{tab:generation_cost} reports expenditure and video volume
for the main-comparison outputs. Final denotes completed runtime;
Generated includes discarded or regenerated clips. The reported
$\overline{G/L}$ averages per-output ratios. Figure~\ref{fig:cost_quality}
shows generation expenditure against RF. The status of outer-loop search
accounting is described in Appendix~\ref{sec:rsi_details}.
\begin{table}[t]
\centering
\caption{Generation expenditure and video volume.}
\label{tab:generation_cost}
\begingroup
\footnotesize
\setlength{\tabcolsep}{0pt}
\setlength{\arrayrulewidth}{0.3pt}
\renewcommand{\arraystretch}{1.10}
\setlength{\aboverulesep}{1.5pt}
\setlength{\belowrulesep}{1.5pt}
\arrayrulecolor{black}
\begin{tabular}{@{}>{\raggedright\arraybackslash\hspace{2pt}}p{0.230\linewidth}|>{\centering\arraybackslash}p{0.140\linewidth}>{\centering\arraybackslash}p{0.140\linewidth}>{\centering\arraybackslash}p{0.140\linewidth}|>{\centering\arraybackslash}p{0.130\linewidth}>{\centering\arraybackslash}p{0.210\linewidth}@{}}
\toprule[0.8pt]
Method & Final (s) & Generated (s) & $\overline{G/L}$ & Cost (\$) $\downarrow$ & DP $\uparrow$ \\
\specialrule{0.4pt}{1.5pt}{1.2pt}
\multicolumn{6}{c}{\textit{FilmEval}} \\
\midrule[0.4pt]
Direct LLM & 397.7 & 547.7 & 1.361 & 9.04 & 0.0\% (0/15) \\
ViMax & 247.7 & 339.2 & 1.362 & \underline{5.92} & 0.0\% (0/15) \\
VideoClaw & 291.0 & 392.2 & 1.342 & 6.69 & 0.0\% (0/15) \\
MovieAgent & 319.5 & 436.8 & 1.348 & 7.36 & 0.0\% (0/15) \\
MM-StoryAgent & 393.0 & 516.8 & 1.316 & 8.46 & 0.0\% (0/15) \\
VGoT & 312.1 & 426.3 & 1.364 & 7.16 & 0.0\% (0/15) \\
\rowcolor{MainTableOurs}
ReVi & 121.8 & 147.5 & 1.214 & \textbf{2.85} & \textbf{100.0\% (15/15)} \\
\specialrule{0.4pt}{1.5pt}{1.2pt}
\specialrule{0.4pt}{0pt}{1.5pt}
\multicolumn{6}{c}{\textit{Recap32}} \\
\midrule[0.4pt]
Direct LLM & 537.6 & 734.8 & 1.360 & 11.68 & 0.0\% (0/12) \\
ViMax & 396.2 & 540.0 & 1.358 & \underline{8.76} & 0.0\% (0/12) \\
MovieAgent & 524.4 & 733.7 & 1.398 & 11.36 & 0.0\% (0/12) \\
\rowcolor{MainTableOurs}
ReVi & 179.3 & 210.3 & 1.172 & \textbf{3.87} & \textbf{100.0\% (12/12)} \\
\specialrule{0.4pt}{1.5pt}{1.2pt}
\specialrule{0.8pt}{0pt}{0pt}
\end{tabular}
\par\vspace{4pt}
\begin{minipage}{\linewidth}
\footnotesize
Main-comparison generation expenditure and video volume. Final and Generated are mean seconds; $\overline{G/L}$ averages per-output ratios. Costs exclude offline search. Rankings apply to cost and DP within each dataset. Bold and underline mark the best and the second-best results, respectively; ties share a rank.
\end{minipage}
\endgroup
\end{table}

\FloatBarrier
\subsection{Human Evaluation and Metric Alignment}
\label{sec:human_alignment}
\paragraph{Study set and presentation.}
The human study uses FilmEval sources C1, C3, C6, C8, C11, and C13:
two from each difficulty tier. For each source, raters evaluate one
completed recap from each of the seven methods in the main comparison.
Method identities are hidden, presentation order is randomized, and
raters have access to the source text and common event reference for
each rating. Three raters score every source--method pair, giving
42 videos, 126 rating records, and 378 dimension scores.

\paragraph{Rating dimensions and aggregation.}
The rating records contain three separate integer scores on a 0--5 scale,
with higher values indicating better quality. Recap Fidelity (RF)
concerns whether the recap preserves important source events and their
causal relations without introducing unsupported developments.
Narrative Pacing (NP) concerns the clarity and progression of the
condensed story, including transitions and narration pace.
Audio-Visual Grounding (AVG) concerns whether the displayed
shots support or complement the spoken account. A video may therefore
score differently across the three dimensions. We first average the
three raters' scores for each video, then give the six sources equal
weight when computing each method's mean. Table~\ref{tab:human_results}
reports these means and the sample SD of the 18 raw ratings per method.

\begin{figure}[!htbp]
\centering
\includegraphics[width=\linewidth]{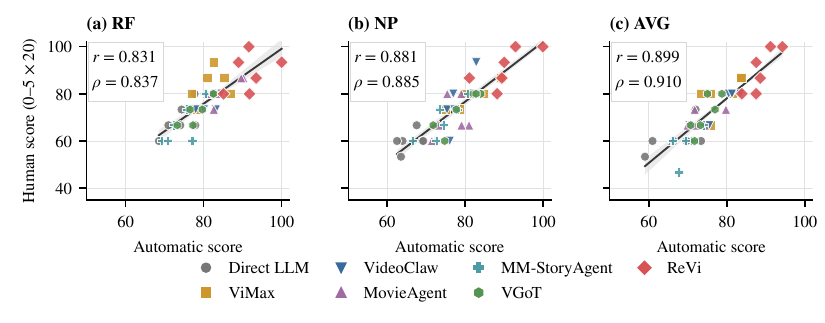}
\caption{\textbf{Automatic scores correlate positively with human
judgments.} Results cover 42 videos from seven methods. Panels show
Recap Fidelity (RF),
Narrative Pacing (NP), and Audio-Visual Grounding (AVG); human means are
scaled from 0--5 by 20. Lines and gray bands show linear fits and
pointwise 95\% source-cluster bootstrap intervals. Insets report
Pearson $r$ and Spearman $\rho$.}
\label{fig:human_evaluation}
\end{figure}

Figure~\ref{fig:human_evaluation} pairs each automatic score with the
same video's mean human rating, scaled by 20. The correlations pool
within-method variation and between-method differences.
Table~\ref{tab:human_agreement} reports source-clustered intervals
and inter-rater agreement. Pooled Pearson correlations for RF/NP/AVG
are 0.831/0.881/0.899, while within-method centered correlations are
0.614/0.688/0.694. Thus, the RF association is not solely due to
differences between method means. Ordinal inter-rater $\alpha$ is
0.462/0.535/0.624. For RF, the three raters show partial agreement,
and ReVi's source-paired human advantage over ViMax is 0.28
[0.06, 0.50] points (Table~\ref{tab:human_paired}). Together these
results offer limited human corroboration of the automatic RF trend;
they do not make automatic and individual human judgments
interchangeable. The six-source sample and pooled correlation design
limit the strength of the claim.

Table~\ref{tab:human_paired} reports ReVi-minus-baseline rating
differences with paired intervals on the original 0--5 scale.
\begin{table}[t]
\centering
\caption{Source-paired human-rating differences.}
\label{tab:human_paired}
\begingroup
\footnotesize
\setlength{\tabcolsep}{0pt}
\setlength{\arrayrulewidth}{0.3pt}
\renewcommand{\arraystretch}{1.10}
\setlength{\aboverulesep}{1.5pt}
\setlength{\belowrulesep}{1.5pt}
\arrayrulecolor{black}
\begin{tabular}{@{}>{\raggedright\arraybackslash\hspace{2pt}}p{0.220\linewidth}|>{\centering\arraybackslash}p{0.255\linewidth}>{\centering\arraybackslash}p{0.255\linewidth}>{\centering\arraybackslash}p{0.255\linewidth}@{}}
\toprule[0.8pt]
Baseline & RF & NP & AVG \\
\specialrule{0.4pt}{1.5pt}{1.2pt}
Direct LLM & +1.00 [+0.67, +1.33] & +1.56 [+1.33, +1.78] & +1.44 [+1.11, +1.78] \\
ViMax & +0.28 [+0.06, +0.50] & +0.61 [+0.33, +0.89] & +0.50 [+0.11, +0.89] \\
VideoClaw & +0.67 [+0.44, +0.89] & +0.72 [+0.33, +1.11] & +0.72 [+0.50, +0.94] \\
MovieAgent & +0.39 [+0.06, +0.72] & +1.06 [+0.94, +1.17] & +1.11 [+0.89, +1.33] \\
MM-StoryAgent & +1.11 [+0.78, +1.44] & +1.11 [+0.72, +1.50] & +1.56 [+1.22, +1.89] \\
VGoT & +0.78 [+0.50, +1.06] & +0.89 [+0.56, +1.22] & +0.94 [+0.50, +1.39] \\
\specialrule{0.4pt}{1.5pt}{1.2pt}
\specialrule{0.8pt}{0pt}{0pt}
\end{tabular}
\par\vspace{4pt}
\begin{minipage}{\linewidth}
\footnotesize
ReVi-minus-baseline human-rating differences and paired intervals on the original 0--5 scale, using six source clusters. Only methods with observed ratings are included.
\end{minipage}
\endgroup
\end{table}
\begin{table}[t]
\centering
\caption{Automatic--human metric agreement.}
\label{tab:human_agreement}
\begingroup
\footnotesize
\setlength{\tabcolsep}{0pt}
\setlength{\arrayrulewidth}{0.3pt}
\renewcommand{\arraystretch}{1.10}
\setlength{\aboverulesep}{1.5pt}
\setlength{\belowrulesep}{1.5pt}
\arrayrulecolor{black}
\begin{tabular}{@{}>{\raggedright\arraybackslash\hspace{2pt}}p{0.090\linewidth}|>{\centering\arraybackslash}p{0.250\linewidth}>{\centering\arraybackslash}p{0.260\linewidth}>{\centering\arraybackslash}p{0.250\linewidth}>{\centering\arraybackslash}p{0.140\linewidth}@{}}
\toprule[0.8pt]
Metric & Pooled $r$ [95\% CI] & Within-method $r$ [95\% CI] & Pooled $\rho$ [95\% CI] & Ordinal $\alpha$ \\
\specialrule{0.4pt}{1.5pt}{1.2pt}
RF & 0.831 [0.819, 0.854] & 0.614 [0.401, 0.793] & 0.837 [0.782, 0.870] & 0.462 \\
NP & 0.881 [0.825, 0.919] & 0.688 [0.548, 0.770] & 0.885 [0.794, 0.932] & 0.535 \\
AVG & 0.899 [0.875, 0.921] & 0.694 [0.615, 0.777] & 0.910 [0.872, 0.929] & 0.624 \\
\specialrule{0.4pt}{1.5pt}{1.2pt}
\specialrule{0.8pt}{0pt}{0pt}
\end{tabular}
\par\vspace{4pt}
\begin{minipage}{\linewidth}
\footnotesize
Automatic--human association on the 42 rated videos. Within-method $r$ subtracts each method's mean before pooling. Correlation intervals resample the 6 source clusters within difficulty, keeping all 7 methods together and re-centering within each replicate. Ordinal alpha uses the three raw ratings per video.
\end{minipage}
\endgroup
\end{table}
\FloatBarrier

\subsection{Difficulty and Source-Paired Uncertainty}
Table~\ref{tab:results_by_layer} reports all three FilmEval tiers,
with five completed outputs per method and tier.
Comparing methods within each tier shows that ReVi retains its lead as
narrative complexity rises. Fidelity generally falls at higher
difficulty: denser plots introduce more event dependencies that must
survive compression, and a brief setup may explain a later action.
The Paper Editor's causal selection targets this pressure. Visual
grounding varies less systematically across tiers, consistent with the
AV Planner's separate task of aligning narration and shots.
Table~\ref{tab:main_paired} reports ReVi-minus-baseline quality
differences and pointwise source-paired interval bounds.
Higher average scores do not imply improvement on every source.
\begin{table}[t]
\centering
\caption{Recap quality by narrative difficulty.}
\label{tab:results_by_layer}
\begingroup
\footnotesize
\setlength{\tabcolsep}{0pt}
\setlength{\arrayrulewidth}{0.3pt}
\renewcommand{\arraystretch}{1.10}
\setlength{\aboverulesep}{1.5pt}
\setlength{\belowrulesep}{1.5pt}
\arrayrulecolor{black}
\begin{tabular}{@{}>{\raggedright\arraybackslash\hspace{2pt}}p{0.105\linewidth}>{\raggedright\arraybackslash\hspace{2pt}}p{0.220\linewidth}|>{\centering\arraybackslash}p{0.110\linewidth}>{\centering\arraybackslash}p{0.110\linewidth}>{\centering\arraybackslash}p{0.110\linewidth}|>{\centering\arraybackslash}p{0.130\linewidth}>{\centering\arraybackslash}p{0.205\linewidth}@{}}
\toprule[0.8pt]
Tier & Method & RF $\uparrow$ & NP $\uparrow$ & AVG $\uparrow$ & DE $\downarrow$ & DP $\uparrow$ \\
\specialrule{0.4pt}{1.5pt}{1.2pt}
Easy & Direct LLM & 76.81 & 67.84 & 64.57 & 199.06 & 0.0\% (0/5) \\
Easy & ViMax & 84.12 & 78.44 & 75.37 & \underline{91.92} & 0.0\% (0/5) \\
Easy & VideoClaw & 81.99 & 80.89 & \underline{79.54} & 128.44 & 0.0\% (0/5) \\
Easy & MovieAgent & \underline{84.91} & \underline{81.15} & 76.94 & 124.62 & 0.0\% (0/5) \\
Easy & MM-StoryAgent & 75.95 & 76.11 & 69.02 & 162.61 & 0.0\% (0/5) \\
Easy & VGoT & 78.91 & 77.49 & 73.92 & 148.39 & 0.0\% (0/5) \\
\rowcolor{MainTableOurs}
Easy & \textbf{ReVi} & \textbf{96.14} & \textbf{93.34} & \textbf{89.39} & \textbf{2.01} & \textbf{100.0\% (5/5)} \\
\specialrule{0.4pt}{1.5pt}{1.5pt}
Medium & Direct LLM & 73.30 & 69.16 & 60.46 & 189.25 & 0.0\% (0/5) \\
Medium & ViMax & 83.39 & 78.86 & \underline{81.78} & \underline{103.05} & 0.0\% (0/5) \\
Medium & VideoClaw & 80.40 & \underline{80.73} & 80.57 & 157.32 & 0.0\% (0/5) \\
Medium & MovieAgent & \underline{84.04} & 78.30 & 73.20 & 160.51 & 0.0\% (0/5) \\
Medium & MM-StoryAgent & 76.65 & 73.88 & 68.81 & 235.69 & 0.0\% (0/5) \\
Medium & VGoT & 78.93 & 76.09 & 72.86 & 159.35 & 0.0\% (0/5) \\
\rowcolor{MainTableOurs}
Medium & \textbf{ReVi} & \textbf{90.97} & \textbf{90.12} & \textbf{86.74} & \textbf{2.34} & \textbf{100.0\% (5/5)} \\
\specialrule{0.4pt}{1.5pt}{1.5pt}
Hard & Direct LLM & 73.28 & 62.54 & 59.84 & 291.78 & 0.0\% (0/5) \\
Hard & ViMax & 78.75 & \underline{78.02} & \underline{78.85} & \underline{131.16} & 0.0\% (0/5) \\
Hard & VideoClaw & 80.94 & 75.99 & 71.79 & 149.97 & 0.0\% (0/5) \\
Hard & MovieAgent & \underline{82.22} & 76.39 & 74.04 & 195.89 & 0.0\% (0/5) \\
Hard & MM-StoryAgent & 74.11 & 71.33 & 71.88 & 259.91 & 0.0\% (0/5) \\
Hard & VGoT & 74.59 & 75.03 & 71.17 & 182.85 & 0.0\% (0/5) \\
\rowcolor{MainTableOurs}
Hard & \textbf{ReVi} & \textbf{87.64} & \textbf{87.15} & \textbf{89.88} & \textbf{2.75} & \textbf{100.0\% (5/5)} \\
\specialrule{0.4pt}{1.5pt}{1.2pt}
\specialrule{0.8pt}{0pt}{0pt}
\end{tabular}
\par\vspace{4pt}
\begin{minipage}{\linewidth}
\footnotesize
FilmEval results by difficulty; all five outputs per method and tier completed. Quality is on a 0--100 scale; duration statistics are percentages. Rankings are within each tier. Bold and underline mark the best and the second-best results, respectively; ties share a rank.
\end{minipage}
\endgroup
\end{table}
\begin{table}[t]
\centering
\caption{Source-paired quality gains over baselines.}
\label{tab:main_paired}
\begingroup
\footnotesize
\setlength{\tabcolsep}{0pt}
\setlength{\arrayrulewidth}{0.3pt}
\renewcommand{\arraystretch}{1.10}
\setlength{\aboverulesep}{1.5pt}
\setlength{\belowrulesep}{1.5pt}
\arrayrulecolor{black}
\begin{tabular}{@{}>{\raggedright\arraybackslash\hspace{2pt}}p{0.220\linewidth}|>{\centering\arraybackslash}p{0.255\linewidth}>{\centering\arraybackslash}p{0.255\linewidth}>{\centering\arraybackslash}p{0.255\linewidth}@{}}
\toprule[0.8pt]
Baseline & RF & NP & AVG \\
\specialrule{0.4pt}{1.5pt}{1.2pt}
\multicolumn{4}{c}{\textit{FilmEval}} \\
\midrule[0.4pt]
Direct LLM & \shortstack[r]{+17.12\\{[+14.52, +19.84]}} & \shortstack[r]{+23.69\\{[+21.33, +25.94]}} & \shortstack[r]{+27.04\\{[+23.45, +30.81]}} \\
ViMax & \shortstack[r]{+9.50\\{[+6.87, +12.27]}} & \shortstack[r]{+11.76\\{[+8.87, +14.70]}} & \shortstack[r]{+10.00\\{[+7.34, +12.57]}} \\
VideoClaw & \shortstack[r]{+10.47\\{[+8.85, +12.10]}} & \shortstack[r]{+11.00\\{[+8.03, +14.17]}} & \shortstack[r]{+11.37\\{[+8.21, +14.73]}} \\
MovieAgent & \shortstack[r]{+7.86\\{[+5.28, +10.48]}} & \shortstack[r]{+11.59\\{[+8.79, +14.23]}} & \shortstack[r]{+13.94\\{[+10.06, +17.59]}} \\
MM-StoryAgent & \shortstack[r]{+16.01\\{[+13.52, +18.50]}} & \shortstack[r]{+16.43\\{[+13.06, +19.85]}} & \shortstack[r]{+18.76\\{[+15.95, +21.68]}} \\
VGoT & \shortstack[r]{+14.10\\{[+11.62, +16.44]}} & \shortstack[r]{+14.00\\{[+11.58, +16.36]}} & \shortstack[r]{+16.02\\{[+12.73, +19.26]}} \\
\specialrule{0.4pt}{1.5pt}{1.2pt}
\specialrule{0.4pt}{0pt}{1.5pt}
\multicolumn{4}{c}{\textit{Recap32}} \\
\midrule[0.4pt]
Direct LLM & \shortstack[r]{+21.74\\{[+19.58, +23.75]}} & \shortstack[r]{+21.85\\{[+19.80, +23.72]}} & \shortstack[r]{+19.63\\{[+13.76, +24.93]}} \\
ViMax & \shortstack[r]{+12.39\\{[+8.95, +15.87]}} & \shortstack[r]{+11.89\\{[+9.68, +14.40]}} & \shortstack[r]{+8.82\\{[+3.94, +13.70]}} \\
MovieAgent & \shortstack[r]{+9.70\\{[+6.94, +12.20]}} & \shortstack[r]{+10.74\\{[+7.66, +13.96]}} & \shortstack[r]{+10.58\\{[+6.36, +14.66]}} \\
\specialrule{0.4pt}{1.5pt}{1.2pt}
\specialrule{0.8pt}{0pt}{0pt}
\end{tabular}
\par\vspace{4pt}
\begin{minipage}{\linewidth}
\footnotesize
Source-paired ReVi gains (points), with interval bounds on the second line of each cell. FilmEval resamples within difficulty; Recap32 resamples its 12 source IDs.
\end{minipage}
\endgroup
\end{table}
\FloatBarrier

\paragraph{Duration and generation expenditure.}
Table~\ref{tab:main_other_intervals} complements the quality comparisons
with duration-error and generation-cost intervals.
\begin{table}[t]
\centering
\caption{Duration-error and cost intervals.}
\label{tab:main_other_intervals}
\begingroup
\footnotesize
\setlength{\tabcolsep}{0pt}
\setlength{\arrayrulewidth}{0.3pt}
\renewcommand{\arraystretch}{1.10}
\setlength{\aboverulesep}{1.5pt}
\setlength{\belowrulesep}{1.5pt}
\arrayrulecolor{black}
\begin{tabular}{@{}>{\raggedright\arraybackslash\hspace{2pt}}p{0.300\linewidth}|>{\centering\arraybackslash}p{0.345\linewidth}>{\centering\arraybackslash}p{0.345\linewidth}@{}}
\toprule[0.8pt]
Baseline & DE (pp) & Generation cost (\$) \\
\specialrule{0.4pt}{1.5pt}{1.2pt}
\multicolumn{3}{c}{\textit{FilmEval}} \\
\midrule[0.4pt]
Direct LLM & [-243.75, -194.99] & [-6.75, -5.45] \\
ViMax & [-121.66, -88.87] & [-3.44, -2.65] \\
VideoClaw & [-159.40, -120.99] & [-4.41, -3.10] \\
MovieAgent & [-167.09, -149.35] & [-4.75, -4.28] \\
MM-StoryAgent & [-224.66, -208.31] & [-5.90, -5.32] \\
VGoT & [-181.71, -135.14] & [-4.96, -3.36] \\
\specialrule{0.4pt}{1.5pt}{1.2pt}
\specialrule{0.4pt}{0pt}{1.5pt}
\multicolumn{3}{c}{\textit{Recap32}} \\
\midrule[0.4pt]
Direct LLM & [-270.60, -148.67] & [-9.90, -5.42] \\
ViMax & [-138.77, -103.24] & [-5.67, -3.90] \\
MovieAgent & [-200.50, -178.49] & [-7.99, -6.94] \\
\specialrule{0.4pt}{1.5pt}{1.2pt}
\specialrule{0.8pt}{0pt}{0pt}
\end{tabular}
\par\vspace{4pt}
\begin{minipage}{\linewidth}
\footnotesize
Paired interval bounds for ReVi minus each baseline on duration error and generation cost. Quality differences are reported in Table~\ref{tab:main_paired}.
\end{minipage}
\endgroup
\end{table}

\paragraph{Editing mechanisms.}
Table~\ref{tab:mechanism_intervals} collects paired quality and
duration-error bounds for the editing ablations.
\begin{table}[t]
\centering
\caption{Intervals for inner-loop component ablations.}
\label{tab:mechanism_intervals}
\begingroup
\footnotesize
\setlength{\tabcolsep}{0pt}
\setlength{\arrayrulewidth}{0.3pt}
\renewcommand{\arraystretch}{1.10}
\setlength{\aboverulesep}{1.5pt}
\setlength{\belowrulesep}{1.5pt}
\arrayrulecolor{black}
\begin{tabular}{@{}>{\raggedright\arraybackslash\hspace{2pt}}p{0.310\linewidth}|>{\centering\arraybackslash}p{0.165\linewidth}>{\centering\arraybackslash}p{0.165\linewidth}>{\centering\arraybackslash}p{0.165\linewidth}|>{\centering\arraybackslash}p{0.185\linewidth}@{}}
\toprule[0.8pt]
Comparison & RF & NP & AVG & DE (pp) \\
\specialrule{0.4pt}{1.5pt}{1.2pt}
\multicolumn{5}{c}{\textit{Inner Loop: ablated minus Full ReVi}} \\
\midrule[0.4pt]
Causal Editing & [-6.37, -3.15] & [-2.50, -0.70] & [-1.94, 0.87] & [0.10, 0.53] \\
TTS Feedback & [-1.04, 0.67] & [-3.82, -1.80] & [-3.19, -1.02] & [6.13, 7.45] \\
Narrative Allocation & [-2.60, -1.45] & [-6.11, -4.60] & [-3.41, -1.89] & [1.69, 1.97] \\
Beat Planning & [-1.92, -0.24] & [-3.21, -1.36] & [-10.76, -7.10] & [-0.13, 0.82] \\
Repair Routing & [-6.73, -2.54] & [-3.38, -2.83] & [-3.36, -2.98] & [1.29, 1.87] \\
\specialrule{0.4pt}{1.5pt}{1.2pt}
\specialrule{0.8pt}{0pt}{0pt}
\end{tabular}
\par\vspace{4pt}
\begin{minipage}{\linewidth}
\footnotesize
Paired interval bounds for inner-loop component ablations. Quality differences are in points; duration-error differences are in percentage points. Every comparison uses six source pairs.
\end{minipage}
\endgroup
\end{table}
\FloatBarrier

\end{document}